\PassOptionsToPackage{pdfpagelabels=false}{hyperref}
\documentclass[useAMS,usenatbib]{mnras}
\pdfoutput=1
\usepackage[pdftex]{graphicx}
\usepackage{url}
\usepackage{amsmath}
\usepackage{natbib}
\usepackage{geometry}
\usepackage{txfonts}
\usepackage{times}

\def \RVIR{\rm{R}_{\rm{200c}}}

\title[Cold Clouds in the CGM of TNG50 MW-like galaxies]{The Circumgalactic Medium of Milky Way-like Galaxies in the TNG50 Simulation -- II: Cold, Dense Gas Clouds and High-Velocity Cloud Analogs}
\author[R. Ramesh, D. Nelson \& A. Pillepich]{Rahul Ramesh$^{1}$\thanks{E-mail: rahul.ramesh@stud.uni-heidelberg.de}, Dylan Nelson$^{1}$ and Annalisa Pillepich$^{2}$
\\
$^{1}$ Universität Heidelberg, Zentrum für Astronomie, Institut für theoretische Astrophysik, Albert-Ueberle-Str. 2, 69120 Heidelberg, Germany\\
$^{2}$ Max-Planck-Institut f\"{u}r Astronomie, K\"{o}nigstuhl 17, 69117 Heidelberg, Germany\\
}

\date{}

\pubyear{2022}

\begin{document}

\maketitle

\begin{abstract}
We use the TNG50 simulation of the IllustrisTNG project to study cold, dense clouds of gas in the circumgalactic media (CGM) of Milky Way-like galaxies. We find that their CGM is typically filled with of order one hundred (thousand) reasonably (marginally) resolved clouds, possible analogs of high-velocity clouds (HVCs). There is a large variation in cloud abundance from galaxy to galaxy, and the physical properties of clouds that we explore -- mass, size, metallicity, pressure, and kinematics -- are also diverse. We quantify the distributions of cloud properties and cloud-background contrasts, providing cosmological inputs for idealized simulations. Clouds characteristically have sub-solar metallicities, diverse shapes, small overdensities ($\chi = n_{\rm cold} / n_{\rm hot} \lesssim 10$), are mostly inflowing, and have sub-virial rotation. At TNG50 resolution, resolved clouds have median masses of $\sim$\,$10^6\,\rm{M_\odot}$ and sizes of $\sim$\,$ 10$\,kpc. Larger clouds are well converged numerically, while the abundance of the smallest clouds increases with resolution, as expected. In TNG50 MW-like haloes, clouds are slightly (severely) under-pressurised relative to their surroundings with respect to total (thermal) pressure, implying that magnetic fields may be important. Clouds are not distributed uniformly throughout the CGM, but are clustered around other clouds, often near baryon-rich satellite galaxies. This suggests that at least some clouds originate from satellites, via direct ram-pressure stripping or otherwise. Finally, we compare with observations of intermediate and high velocity clouds from the real Milky Way halo. TNG50 shows a similar cloud velocity distribution as observations, and predicts a significant population of currently difficult-to-detect low velocity clouds.
\end{abstract}


\section{Introduction}
\label{intro}

The circumgalactic medium (CGM), the halo of gas surrounding galaxies, is believed to be critically linked to their formation and evolution. While the bulk of the CGM is home to a volume-filling diffuse warm-hot gas phase, the CGM is also commonly multi-phase. It can host small clouds of cold, dense gas (see \citealt{donahue2022} for a recent review of the CGM).

Historically observed through their HI emission, compact gas clouds ($\lesssim 1-10$ kpc) are observed in the Milky Way halo at large velocities with respect to the local standard of rest (LSR), and have hence been named high-velocity clouds (HVCs; e.g. \citealt{muller1963, wakker1991, wakker1997}). More recently, such clouds have been further differentiated based on their velocities, into so called low-velocity clouds (LVCs), intermediate-velocity clouds (IVCs) and very high-velocity clouds \citep[VHVCs; e.g.][]{hafner2001,peek2009,lehner2011}. While early studies mainly targeted the Milky Way halo, which is believed to contain many thousands of such clouds \citep[e.g][]{putman2002, moss2013}, more recent explorations have begun identifying clouds around external galaxies as well \citep[e.g][]{gim2021}.

Although these clouds have been observed for many decades now, there remain multiple open questions. For example, their origin is highly debated: while a fraction of these clouds may be related to the stripping of gas as satellites infall into the potential minimum of their host halo \citep[e.g.][]{olano2008}, theory also suggests that clouds can form via condensation of hot halo gas \citep[e.g.][]{binney2009, juong2012, fraternali2015} and `fountain' flows of gas in and around galaxies \citep[e.g.][]{fraternali2006}. Recently, \cite{lehner2022} and \cite{marasco2022}, using a sample of (observed) IVCs and HVCs, showed that both diffuse `rain-like' inflows and collimated outflows are present in their sample, adding weight to the galactic fountain scenario. They, however, do note that only $\sim30$ per cent of their clouds are outflowing, which is slightly lower than expected with the galactic fountain scenario, although such a bias may just stem from the collimated geometry of outflows.

Since HVCs were first observed through their 21cm line emission, it was assumed that these clouds are typically dominated by cold gas, and are pristine with respect to their metallicity content. However, recent observations of metal line absorption along the line of sight to quasars and stars have shown that these clouds may indeed contain non-negligible amounts of metals. For instance, using data collected by the Far Ultraviolet Spectroscopic Explorer (FUSE), \cite{sembach2000} observed OVI absorption in HVCs, while \cite{savage2000} reported the correlation between MgII absorption and known locations of HVCs using the Hubble Space Telescope (HST). \cite{lehner2001} showed the presence of CII, OI, SII, SIII, and SIV along a sightline that intercepts a known HVC. Using observations with the Cosmic Origin Spectrograph (COS), \cite{richter2017} noted the presence of SiIII in HVCs. Further, the metallicity distributions across a sample of HVCs can be large, with values ranging from highly sub-solar to super-solar (e.g. \citealt{wakker2001, fox2016}); on the other hand, the range of metallicities for IVCs is relatively narrower \citep{lehner1999, wakker2001}, possibly indicating varied and different origins. Finally, observations of H$\alpha$ emission from these clouds suggest the presence of slightly warmer gas, potentially in interface layers between the cold clouds and the background hot halo \citep{tufte1998, hafner2001}.

While distances to HVCs are generally unconstrained, given that they are predominantly observed through HI emission, more recent studies have begun providing (upper limit) distance estimates for a large number of clouds. For instance, absorption features along the line of sight of stars have been used to estimate a distance of $\leq 10$ kpc to Complex C \citep{wakker2007}, $\leq 15$ kpc to Complex GCP \citep{wakker2008} and $\sim$4.4 kpc to Complex WD \citep{peek2016}. In addition to absorption studies, measurements of magnetic field strengths around clouds may be useful in estimating their distances \citep{gronnow2017}. A consequence of the (earlier) lack of distance estimates was that basic properties of clouds like their masses and sizes were poorly understood. Current work now suggests that the aforementioned properties of clouds show large diversity, varying from large HI masses of $\sim10^7\, \rm{M_\odot}$ (Complex C, \citealt{thom2008}) down to $\lesssim 10^5\, \rm{M_\odot}$ (\citealt{wakker2001, adams2013}). Despite employing these novel techniques to estimate distances to clouds, an unbiased view of HVCs in the Milky Way is currently not available, such that the typical distance to HVCs is not known, and neither are the mean mass or size or e.g. the mass/size distributions.

Another mystery surrounding these cold CGM clouds is their expected lifetime, i.e. long-term survival in the face of fluid instabilities and mixing processes. A large variety of numerical `cloud crushing' simulations have studied these questions. Early works typically suggested that these clouds are short-lived, either because they are destroyed \citep{klein1994, schneider2017} or broken into smaller fragments \citep{mellema2002} by shocks and/or hot winds driven by supernovae. That is, the cloud-shredding time scale is of order the time required to accelerate to their relatively large velocities \citep{zhang2017}.
However, more recent studies propose that they may actually not be as unstable as previously thought. For instance, the Kelvin-Helmholtz (KH) instability can create a turbulent mixing layer between the cloud and the surrounding halo gas, giving rise to a warm gas interface layer that cools rapidly \citep{nelson2020}, thereby contributing a sizeable amount of cold gas to the cloud \citep{gronke2018, fielding2020}. In particular, \cite{gronke2020} suggest that radiative cooling may be more important than, and win against, the KH instability.

In addition, magnetic fields may play an important role in stabilizing these clouds: either through the associated magnetic pressure that counterbalances the thermal pressure of the ambient hot medium \citep{nelson2020}, through magnetic tension that suppresses buoyant oscillations of the condensing gas \citep{ji2018}, or possibly by enhancing the Rayleigh-Taylor instability that leads to larger rates of condensation \citep{gronnow2022}. Similar to magnetic fields, cosmic rays may also contribute pressure support, thereby providing additional stability to these cold gas clouds (for e.g., \citealt{butsky2018}). Furthermore, under the right conditions, cold clouds of gas may coagulate, i.e., small fragments may coalesce into a larger mass \citep{gronke2022}. 
However, it is to be noted that the outcomes of these simulations generally depend upon the specifics of the clouds vs background properties and setup, the physical processes included, and numerical resolution \citep[e.g.][]{jennings2022}. As of today, cloud survival remains an open topic.

Idealized numerical studies of cloud survival have a fundamental limitation: they are all non-cosmological, `wind tunnel' simulations, i.e they assume the existence of a pre-formed cloud of a given composition, and evolve this cloud in the presence of a hot-ambient medium. As a result, these simulations do not account for the complexity and structure of a realistic CGM, its diversity across the galaxy population, nor its evolution across cosmic time. In addition, such numerical experiments cannot answer questions that are input assumptions: namely, the origin and properties of clouds, including their mass and size distributions, as well as their composition, and relative dynamics with respect to the background CGM.

In this paper, we use the TNG50 simulation of the IllustrisTNG project to investigate the existence and properties of cold clouds around Milky Way (MW)-like galaxies. TNG50 is a cosmological uniform-volume simulation that has been shown to be able to realize small-scale, cold gas structures in the CGM of high-mass elliptical galaxies, as traced by neutral HI and MgII \citep{nelson2020}. As a magneto-hydrodynamical simulation over a large cosmological volume, TNG50 is able to account for the physically plausible, complex, and diverse CGM of MW-like galaxies, together with their embedded cloud populations. Here in particular we focus on the $z=0$ CGM of 132 MW-like galaxies whose global gaseous-halo properties have already been extensively characterized by \citealt{ramesh2022}, focusing on the physical properties of their cold clouds.

This paper is organised as follows: in Section~\ref{methods}, we provide a brief description of the IllustrisTNG simulations and TNG50, the sample selection process, and the algorithm we employ to identify clouds. In Section~\ref{results}, we present the results of this work: the abundance and global statistics of cold clouds in the CGM, including their intrinsic physical properties and distribution through the halo. We demonstrate a suggestive correlation between the location of clouds and satellite galaxies with relatively large baryon fractions. Finally, in Section~\ref{summary}, we summarise our main results and conclude.

\section{Methods}\label{methods}

\subsection{The TNG50 simulation}\label{TNG}

In this paper, we use the TNG50-1 (hereafter, TNG50) simulation \citep{pillepich2019, nelson2019} of the IllustrisTNG project \citep{springel2018, naiman2018, pillepich2018b, marinacci2018, nelson2018}. These are a set of cosmological magneto-hydrodynamical simulations run with the code \textsc{Arepo} \citep{springel2010}. IllustrisTNG builds on its predecessor, the original Illustris simulation \citep{genel2014, vogelsberger2014a, vogelsberger2014b, sijacki2015}, employing a modified model for galaxy physics \citep{weinberger2017, pillepich2018a}, along with the addition of magnetic fields \citep{pakmor2014}, an important physical quantity that was previously absent in Illustris.

TNG50 simulates a volume of $\sim$(50 cMpc)$^3$ at an average baryonic mass resolution of $\sim 8 \times 10^4$ M$_\odot$. It remains the highest resolution cosmological simulation that exists at this volume, and the largest volume cosmological simulation that exists at this resolution. As explained in detail in \cite{pillepich2018a}, the TNG simulations include recipes for a variety of physical processes that are believed to play a critical role in galaxy formation and evolution. We refer the reader to that paper for all details of the TNG model. 

We briefly mention that the TNG simulations employ the \cite{springel2003} two-phase subgrid model to stochastically convert star-forming gas ($n_{\rm H} \geq 0.1$ cm$^{-3}$) to stars. As a result, the temperature of star-forming gas is `effective', i.e. not physical. For our analysis, we always set the temperature of star forming gas to its cold phase value, $10^3$\,K\footnote{This is the temperature invoked in the TNG model for the cold phase of the subgrid ISM model \citep{springel2003}, and does not necessarily reflect the true gas temperature in star-forming gas complexes, clouds, or cores. For our purposes, any value below $T<10^4$K will not significantly impact the present results.}, which dominates by mass $(\geq 90)$ per cent.

While the amount of neutral hydrogen content of gas is directly tracked and output by the simulation, the fraction of this component in atomic hydrogen is not. We use the \cite{gnedin2011} H$_2$ model to estimate the fraction of neutral hydrogen in H$_2$, thereby arriving at an estimate of neutral atomic hydrogen by subtracting the two values, following \cite{popping2019}.

The TNG simulations adopt a cosmology consistent with the Planck 2015 analysis \citep{planck2016}, with: $\Omega_\Lambda = 0.6911$, $\Omega_{\rm m} = 0.3089$, $\Omega_{\rm b} = 0.0486$ and $h = 0.6774$.

\subsection{The Milky Way-like galaxy sample}

In this work, we use the same sample of MW-like galaxies as \cite{ramesh2022}\footnote{This is a subset of the TNG50 `Milky Way/M31 sample' of galaxies presented in \textcolor{blue}{Pillepich et al. (in prep)}, where these galaxies are discussed.}: these are a set of 132 galaxies that are (i) centrals, i.e. they lie at the potential minimum of their host friends-of-friends \citep{davies1985} haloes, (ii) reside in haloes that are not overly massive (virial mass, M$_{\rm{200,c}} < 10^{13}~\rm{M}_\odot$), (iii) have a stellar mass, measured within a 3D aperture of 30kpc, in the range $10^{10.5}\,\rm{M}_\odot$ to $10^{10.9}\,\rm{M}_\odot$, (iv) are reasonably well isolated (no other galaxy having M$_\star > 10^{10.5}~\rm{M}_\odot$ within a distance of $500$\,kpc), and (v) are disky, either through visual inspection of stellar light maps or based on a constraint on the minor-to-major axis ratio of the stellar mass distribution ($s < 0.45$). We refer the reader to {\color{blue} Pillepich et al. in prep} for the motivation behind the choice of these criteria, and the extent to which such a sample captures our real Milky Way.

Following \cite{ramesh2022}, we (i) define the circumgalactic medium (CGM) as the region bounded by $[0.15, 1.0] \times \RVIR$ of the corresponding halo, (ii) exclude gas that is gravitationally bound to satellite galaxies (i.e. all galaxies in a halo that aren't centrals), as identified by the substructure identification algorithm \textsc{subfind} \citep{springel2001}, and (iii) consider all gas in the simulation volume, i.e. we do not restrict our selection of gas cells based on their friends-of-friends halo membership. Unless otherwise stated, the stellar mass of a galaxy is defined as the sum of the mass of all stars within an aperture of 30\,pkpc. Star formation rates are defined by the stellar mass formed within an aperture of 30\,pkpc, over the last billion years \citep{pillepich2019, donnari2019}.

\begin{figure*}
\centering 
\includegraphics[width=16cm]{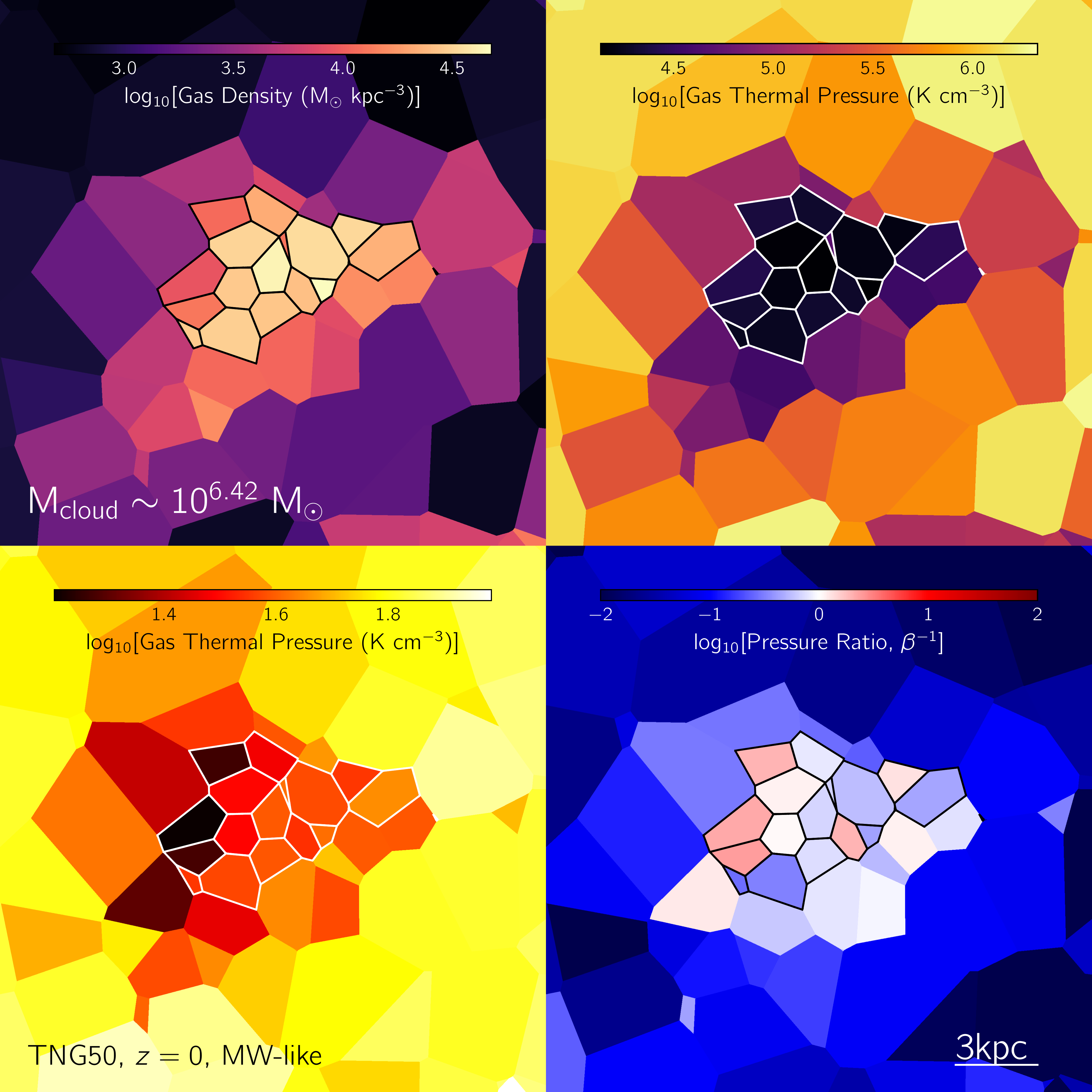}
\caption{A visualisation of the physical properties of a single cloud (M$_{\rm{cloud}} \sim 10^{6.4}$ M$_\odot$) from the CGM of a Milky Way-like galaxy in the TNG50 simulation. Each panel shows a slice of the Voronoi mesh centred on the same cloud. Voronoi cells that are members of the cloud are outlined by either black or white lines. In cyclic order, from the top left to bottom left, cells are colored by their density, temperature, ratio of magnetic to thermal pressure, and thermal pressure. The central region of this cloud is dense and cold, and relative to the local environment, this region is: overdense, cooler, thermally under-pressurized, and has $\beta \sim 1$ with magnetic and thermal pressures in rough balance.}
\label{fig:cloudAlgo2}
\end{figure*}

\subsection{Cold gas cloud identification algorithm}\label{cloud_algo}

A consistent way to identify clouds in simulations run with \textsc{Arepo} is via spatially contiguous sets of Voronoi cells \citep[following][]{nelson2020}. In particular, `natural neighbors' of a Voronoi cell are those that share a face, i.e. they directly touch. An ensemble of naturally connected Voronoi cells is a union of convex polyhedra, and this geometrical structure is well suited to represent clouds of arbitrary size and shape. Our algorithm consists of two steps:

\begin{enumerate}
    \item We first identify `cold' gas as those cells whose temperature $T<10^{4.5}$K. We note that our algorithm is not overly sensitive to this chosen threshold value. For instance, varying the threshold by $\sim 0.2$ dex only changes the average number, and size, of clouds by $\sim 10$ per cent. This suggests that the different temperature cutoffs identify the same clouds, but include different amounts of interface gas around their cores \citep[see][]{nelson2020}.
    
    \item Once cold gas cells are identified, we group them into clouds by looking for contiguous sets of Voronoi cells. For the main results of this work, we consider only those clouds that contain at least ten Voronoi member cells. Various numerical experiments have concluded that clouds need to be resolved by at least few tens to hundreds of resolution elements to adequately capture their growth and evolution \citep[e.g.][]{klein1994, nakamura2006, yirak2010, goldsmith2016, pittard2016}. We therefore consider a lower limit of ten member cells per cloud. This avoids issues with low number statistics, while restricting our analyses to marginally well-resolved clouds. We note that most of our results are \textit{qualitatively} similar even if a slightly lower threshold of member cells per cloud is adopted. Cases that differ qualitatively are shown explicitly in the corresponding panels.
\end{enumerate}

We run the algorithm for all gas within $\RVIR$, with satellite gas excluded.  For each halo, the algorithm typically returns one `massive' cloud of mass $\gtrsim 10^{8.5} \rm{M_\odot}$ that lies close to the centre of the halo, i.e. the galaxy itself. We exclude this object from our analysis. In addition, unless otherwise stated, we exclusively consider clouds that lie outside $0.15 \times \RVIR$, which is our inner boundary for the CGM.

\section{Results}\label{results}

\subsection{\textsc{arepo}, TNG50 and CGM cold clouds}

To provide an idea of the cold clouds that are present in the TNG50 simulation in the CGM of MW-like galaxies, Figure~\ref{fig:cloudAlgo2} shows a visualisation of an individual cloud identified by our algorithm (Section~\ref{cloud_algo}). This cloud has a mass of M$_{\rm{cloud}} \sim 10^{6.4}$ M$_\odot$, and is composed of 33 Voronoi cells. Each panel shows a slice of the Voronoi mesh in a small region centred on the chosen single small cloud. The Voronoi cells that belong to this cloud are outlined with white/black lines. Note that only a subset of the 33 Voronoi cells are visible in these panels, since cells displaced along the direction perpendicular to the screen are not visible in the slice.

In the top left panel, we color by (three-dimensional) gas density: at the centre of the cloud, gas is more dense compared to its outskirts, and the density drops with increasing distance from the cloud center, into the background region surrounding this cloud. Since \textsc{arepo} dynamically (de-)refines the mesh to ensure that the mass of all gas cells is roughly equal, the Voronoi cells at the centre of the cloud are naturally smaller, resulting in higher spatial resolution. 

The top-right panel shows the temperature of gas: the centre of this cloud is dominated by cold gas (T $\lesssim 10^{4.5}$K), and is embedded in a predominantly hot medium (T $\gtrsim 10^{5.5}$K). Although the numerical resolution in TNG50 at these small scales is limited, we see that the cloud and background regions are separated by an intermediate warm-phase ($10^{5.5}$K $\gtrsim$ T $\gtrsim 10^{4.5}$K), which may play a role in increasing the longevity of such clouds (e.g. \citealt{gronke2018,fielding2020,nelson2020,abruzzo2022}).

In the bottom left panel, we color by the thermal pressure of gas. Gas in this cloud is thermally under-pressurized with respect to the background. The distribution of thermal pressure is asymmetric, due to the asymmetric distribution of temperature and density. 

Lastly, in the bottom-right panel, we show the ratio of magnetic to thermal pressures, $\beta^{-1}$. The majority of the cloud has $\beta \sim 1$, indicating that magnetic and thermal pressure are in rough equipartition. While certain regions of this cloud are dominated by magnetic pressure, the thermal component dominates in others. This is in contrast to clouds in much larger group-mass haloes in TNG50, where magnetic pressure strongly dominates within cold clouds, likely due to the higher ambient densities \citep{nelson2020}.

The properties of the TNG50 cloud showcased in Figure~\ref{fig:cloudAlgo2} are typical, as we expand upon in the following sections.

\subsection{Location and number of cold clouds around MW analogs}\label{sec:clouds_basic}

\begin{figure*}
\centering 
\includegraphics[width=16.0cm]{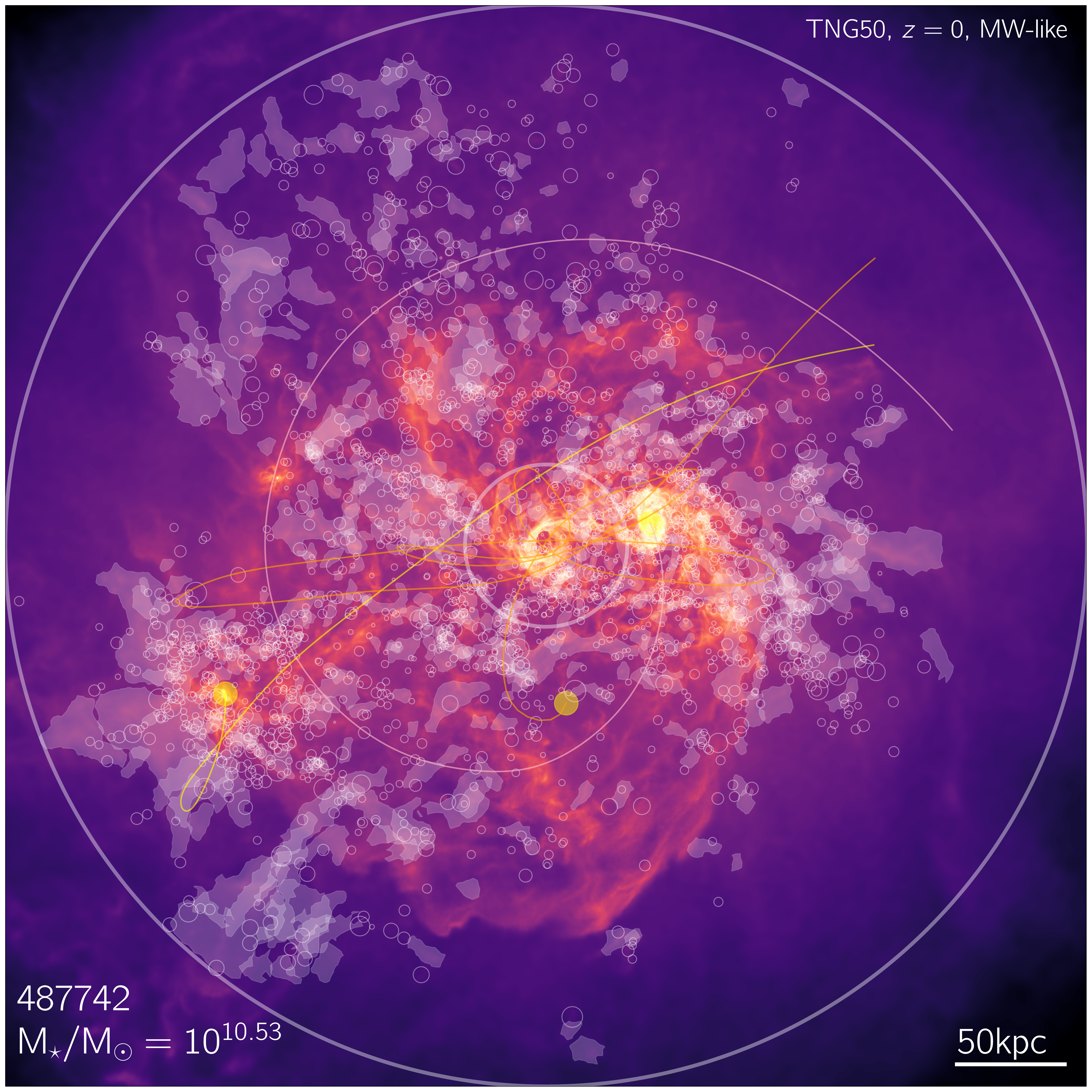}
\includegraphics[width=8.5cm]{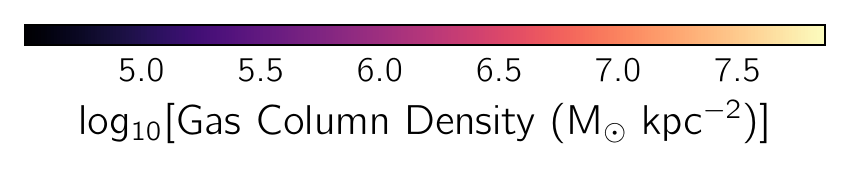}
\includegraphics[width=8.5cm]{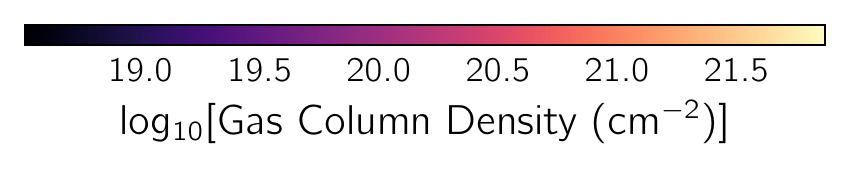}
\caption{A visualisation of the distribution of clouds around one of the TNG50 MW-like galaxies from our sample (subhalo 487742). Background color shows gas column density, while the two concentric white circles are drawn at radii $[0.15, 1.0] \times \RVIR$. Unfilled white circles in the foreground correspond to unresolved clouds, i.e. those composed of less than 10 gas cells and thus not included in our main analyses. Translucent shapes correspond to more massive clouds, which comprises our `fiducial' sample. The yellow circles show the location of satellite galaxies with large baryon ratios ($> 10$ per cent), while the connected curves show the past trajectory of these satellites. Clouds are preferentially seen to lie close to satellites, or close to the their past trajectories. We interpret the former as gas that has been `freshly' stripped, and the latter as gas that was stripped in the past.}
\label{fig:cloudPosSatPos}
\end{figure*}

Zooming out from a single cloud, we visualise the entire distribution of clouds in one halo from our sample of TNG50 MW-like galaxies. Figure~\ref{fig:cloudPosSatPos} shows the gas column density in the background, integrated over an extent of $\pm \RVIR$ along the line of sight direction. The two concentric white circles are drawn at radii $[0.15, 1.0] \times \RVIR$, the chosen inner and outer boundaries of the CGM. In the foreground, the distribution of clouds is shown: clouds composed of less than ten Voronoi member cells (which we do not generally include in our analyses) are drawn as unfilled white circles, with their radii scaled by the size of the cloud; more massive clouds, i.e. the clouds we consider in this work, are shown instead directly by their (projected) convex hulls, as translucent filled shapes, to provide a better sense of their complex structure. Note that regions where shapes appear `brighter' correspond to overlapping clouds along the line of sight. It is clear that (i) an enormous number of cold clouds are present in the CGM of MW-like galaxies, and (ii) CGM cold gas clouds come in a variety of shapes and sizes.

The yellow filled circles correspond to the positions of satellite galaxies with large baryon ratios ($> 10$ per cent; at $z=0$), and the curves connected to each trace their past orbital trajectories. A large number of clouds lie close to these satellites, or close to their past orbits.\footnote{We encourage the reader to explore the significant diversity of cloud populations across the sample in our \href{https://www.tng-project.org/explore/gallery/ramesh23/}{online infinite gallery}.} However, there are some clouds that are close to neither, and such clouds could be linked to satellites that are no longer baryon rich (but were at some point in the past), or could just have drifted away after having been stripped. This figure conveys the concept that we expand upon in the next sections \citep[originally discussed in][]{nelson2020}: clouds cluster around satellites. Satellite galaxies clearly make an important contribution to the cold gas contents of the CGM (\textcolor{blue}{Rohr et al. in prep}). In addition, many clouds also exist close to other clouds, as opposed to being uniformly distributed through the halo: clouds cluster around themselves. Both these correlations are key results of this work, which we later quantify in Section~\ref{sec:clouds_spatial_clustering}. 

\begin{figure}
\centering 
\includegraphics[width=8cm]{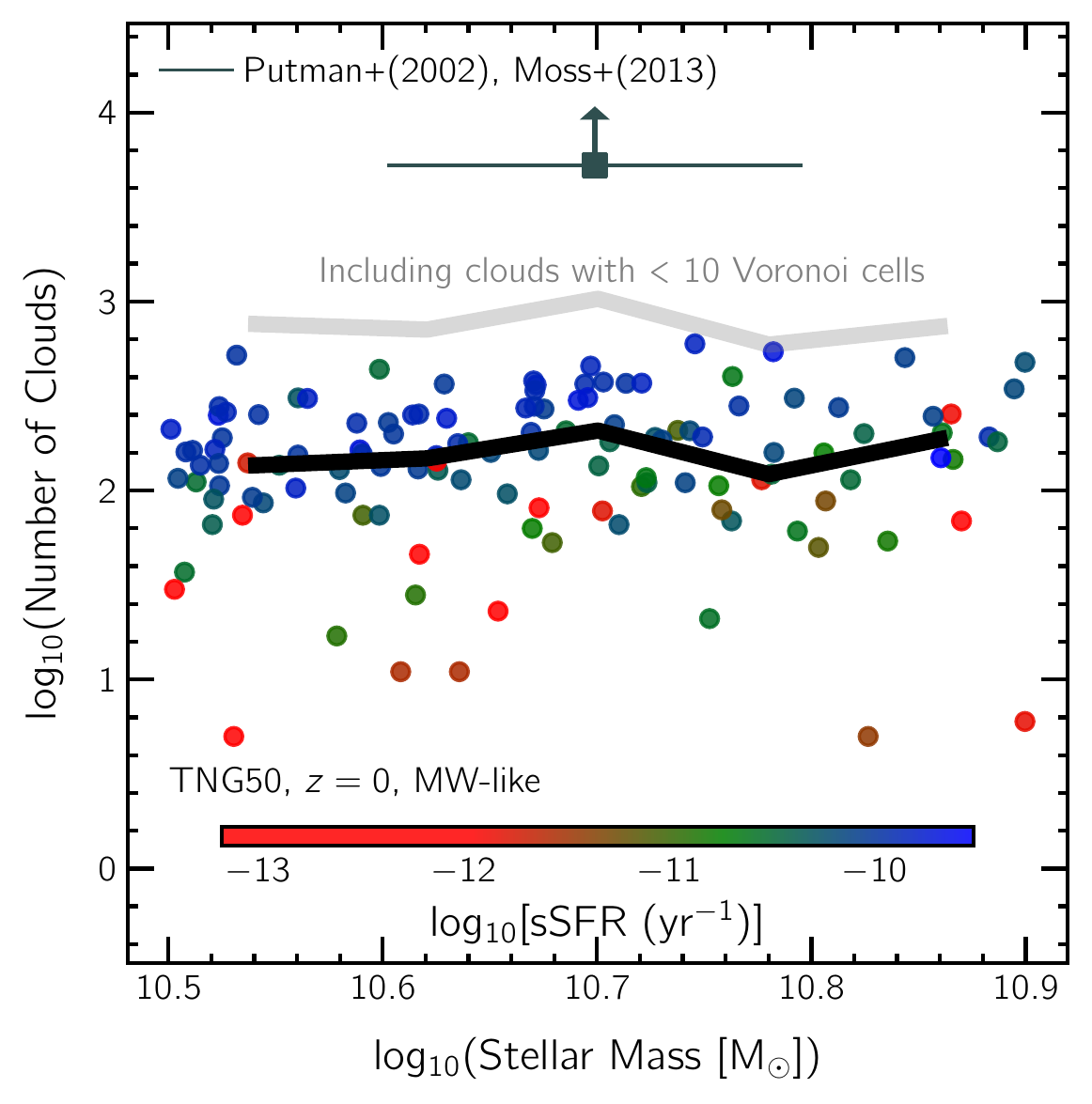}
\caption{Number of clouds in the CGM around each TNG50 MW-like galaxy as a function of its stellar mass, with scatter points colored by the sSFR of the galaxy. On average, the number of clouds show a flat trend with stellar mass of the galaxy, albeit with a large scatter at fixed galaxy stellar mass. A trend in the colors of the points is, however, seen: galaxies with higher sSFR's have a larger number of clouds in their CGMs. The black (gray) solid curve depicts median results for our fiducial (alternative) definition of clouds, i.e. with a minimum number of 10 (1) cells each.}
\label{fig:numClouds}
\end{figure}

As discussed in \cite{ramesh2022}, the properties of halo gas across our sample of MW-like galaxies exhibit a large diversity. To explore the effect of this galaxy-to-galaxy diversity on the properties of clouds, Figure~\ref{fig:numClouds} plots the number of CGM clouds as a function of stellar mass of the galaxy. The circles, colored by the specific star formation rate (sSFR) of the galaxy, denote individual haloes, while the median across galaxies is shown in the black solid curve.

The CGM of TNG50 MW-like galaxies typically contain of order one hundred reasonably resolved clouds, and of order one thousand marginally resolved clouds. The median shows no significant trend as a function of stellar mass, although a large scatter is evident, with numbers varying between a few clouds per halo to a few hundred. However, a strong trend is apparent in the vertical direction across the scatter: galaxies with higher values of sSFR preferentially host a greater number of clouds in their haloes. As discussed extensively in \cite{ramesh2022}, such a trend with properties related to \textit{cold gas} may arise due to two factors:  either (i) because outflows generated by the central super-massive black hole (SMBH) both suppress star formation in the galaxy and heat up CGM gas \citep[e.g.][]{weinberger2017, zinger2020}, or (ii) because of a physical connection between the flow of gas through the CGM and the SF activity of the galaxy. The latter case suggests that these clouds may play a role in sustaining star formation in the galaxy by replenishing the required `fuel', as is commonly proposed for HVCs in the Milky Way halo \citep[e.g.][]{lepine1994}. In the former case, it is possible that high-velocity outflows generated by the kinetic mode of SMBH feedback in TNG50 destroy clouds that are already present \citep[analogous to clouds being shredded by supernova feedback, e.g. ][]{schneider2017}; but it may also be the case that clouds are prevented from forming in the first place.

A robust comparison with observations of the real Milky Way halo is not possible. However, we take a first step with the gray square, which shows a lower limit for the number of observed HVCs. To arrive at this estimate, we stack the catalogs presented in \cite{putman2002} and \cite{moss2013}. These contain a total of 1956 and 1693 clouds respectively, with an overlap of 1021 clouds: the stacked sample therefore contains a total of 2628 unique clouds. Since both these surveys observe only the southern sky, we assume that the northern sky contains an equal number of clouds, and multiply the sample size by two, yielding a lower limit of 5256 clouds. For the stellar mass, we assume a value of $\rm{M_{\rm{\star, MW}}} \sim (5 \pm 1) \times 10^{10}~\rm{M_\odot}$ \citep{bland2016}.

\begin{figure*}
\centering 
\hspace{3cm}
\includegraphics[width=7cm]{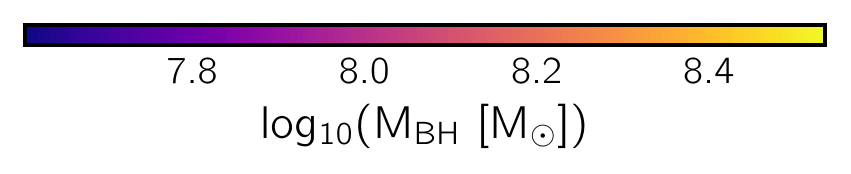}
\newline
\includegraphics[width=8cm]{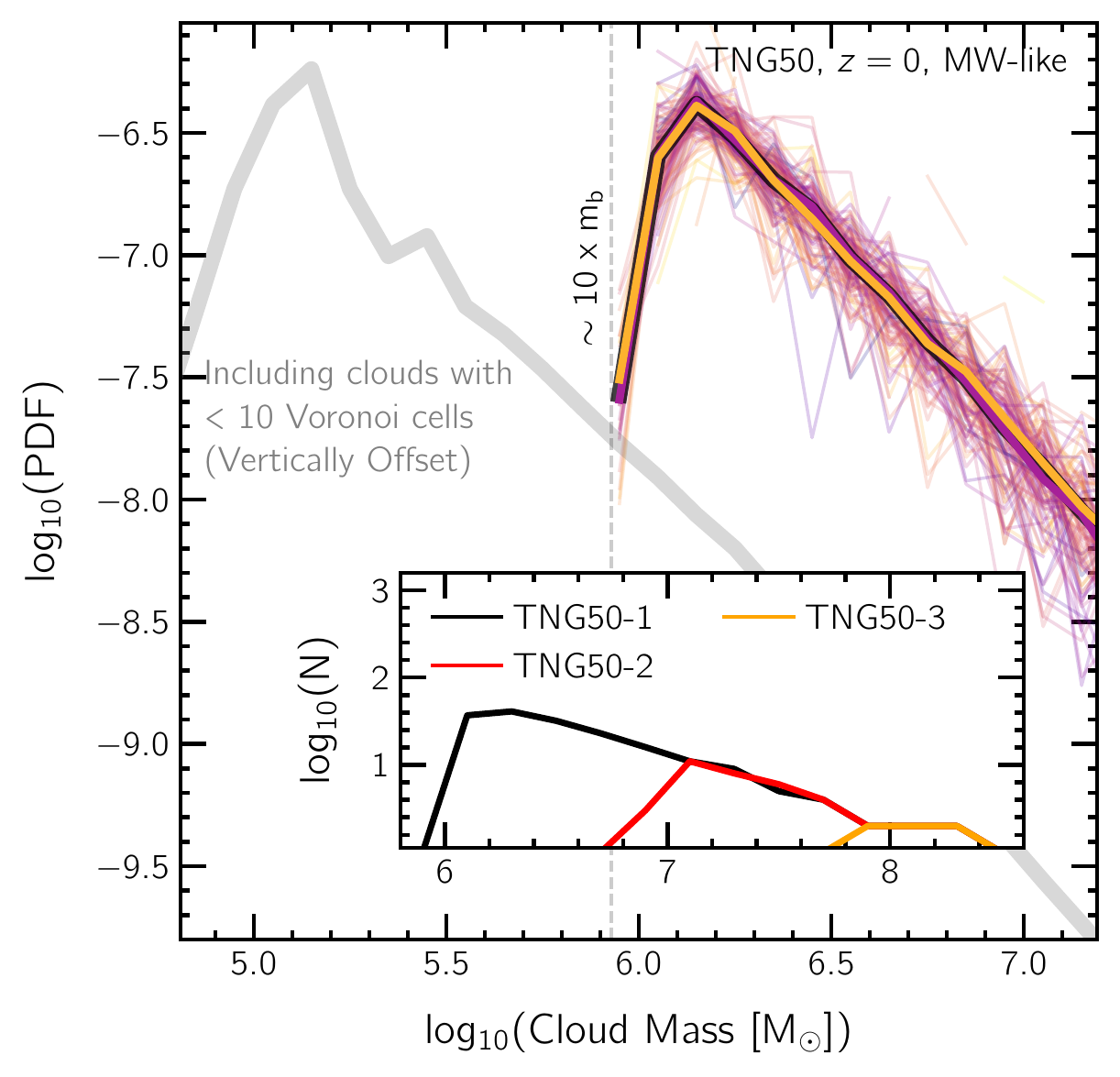}
\includegraphics[width=8cm]{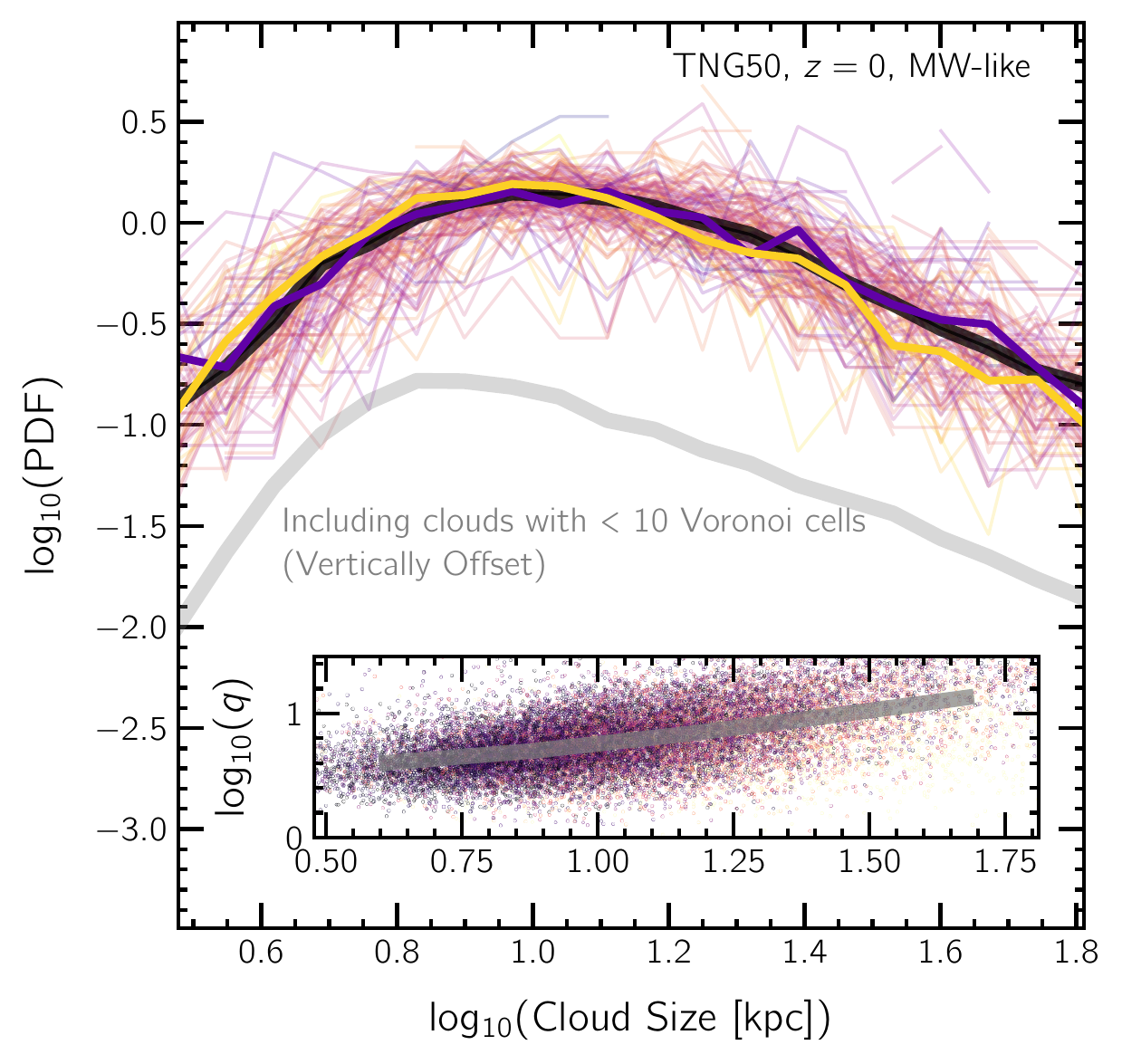}
\includegraphics[width=8cm]{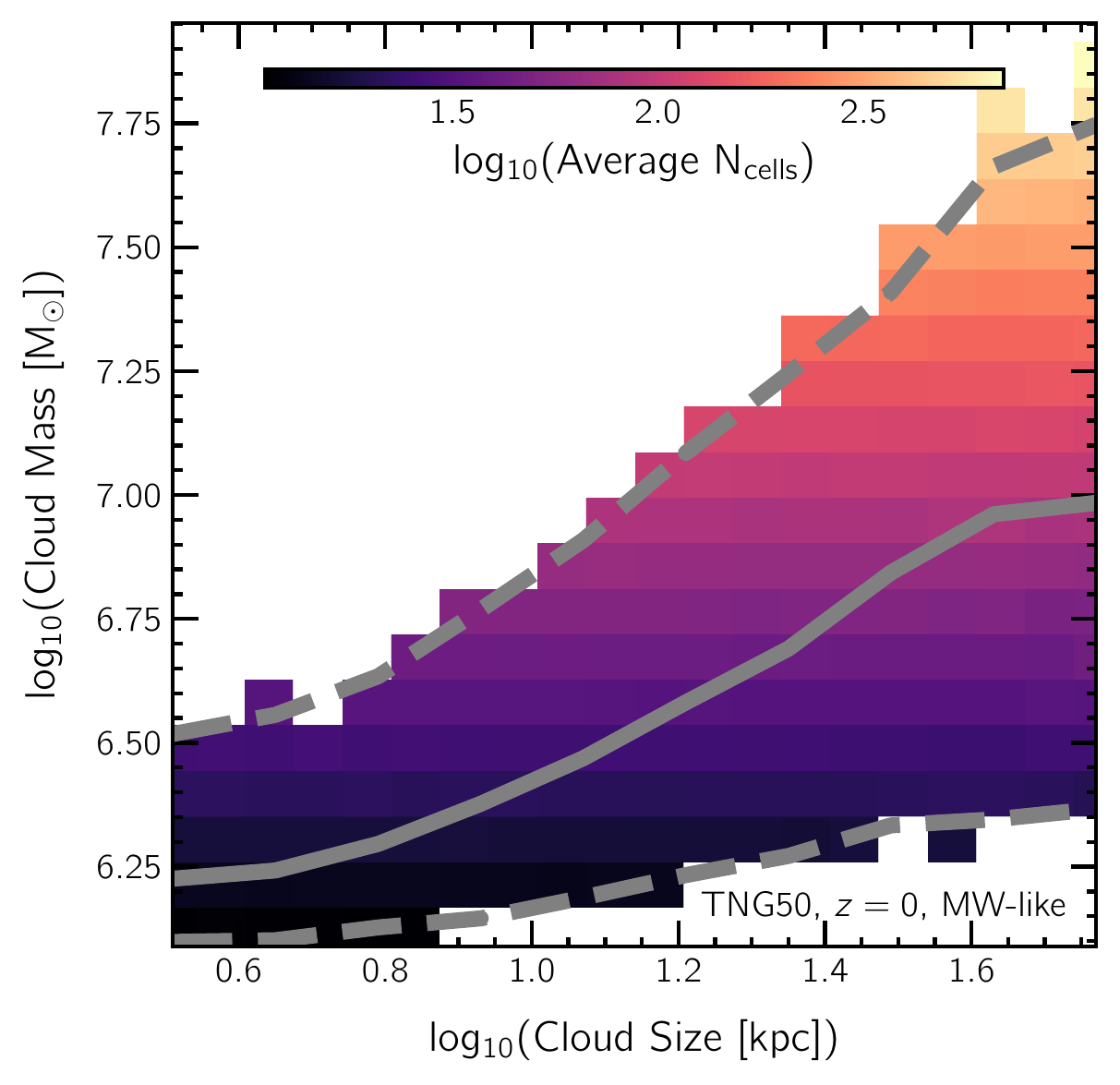}
\includegraphics[width=8cm]{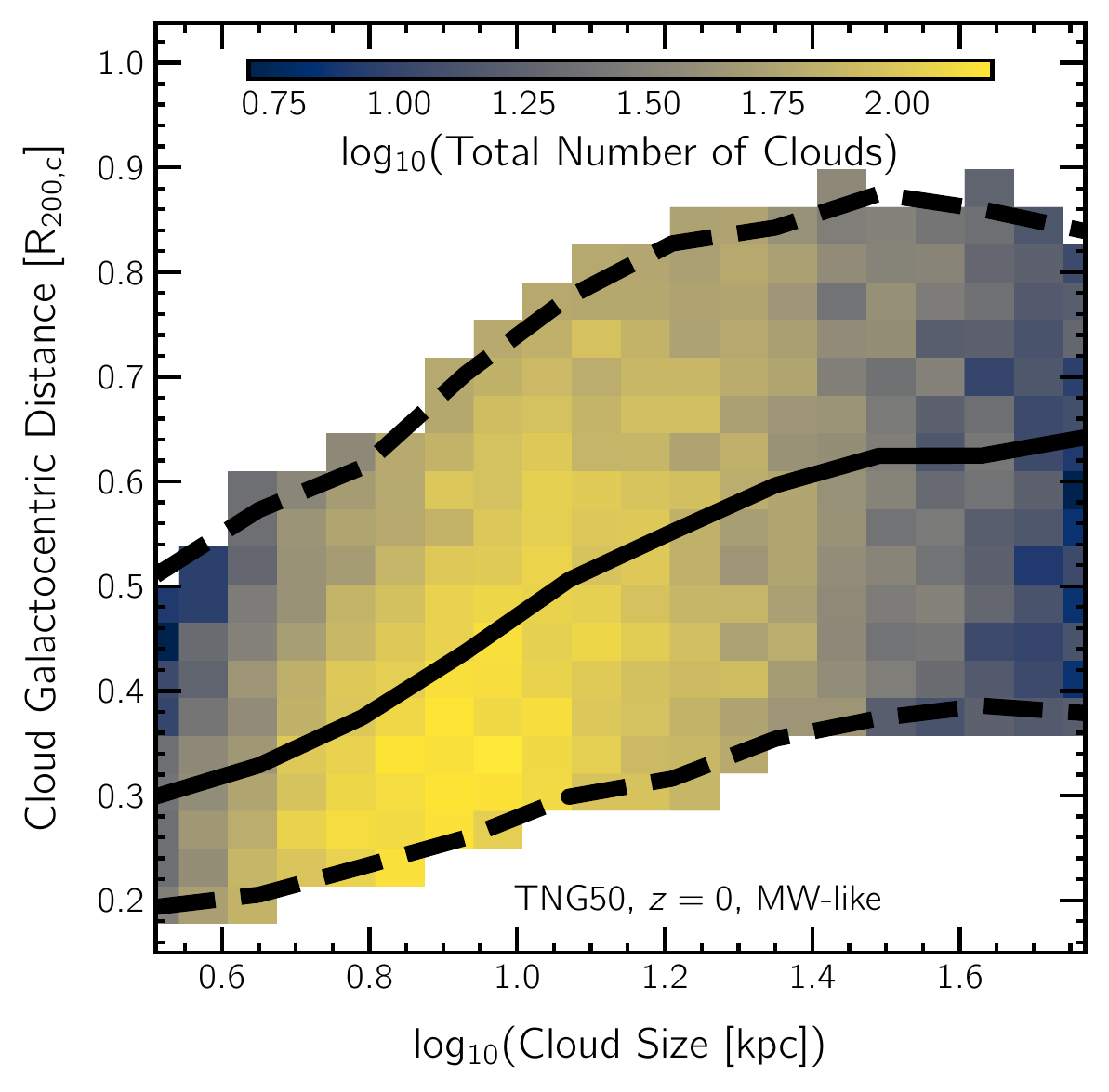}
\caption{Properties of the clouds in the CGM of TNG50 MW-like galaxies: the top-left and top-right panels show the cloud mass and cloud size distributions, respectively. In each panel, the thin curves correspond to individual haloes, color coded by central SMBH mass. Solid black curves show the median PDF across the sample, and purple and yellow curves correspond to median distributions of the two extreme octile regions, splitting by SMBH mass. The top-left panel includes an inset where the mass function of clouds is compared to lower resolution versions (TNG50-2 and TNG50-3) of TNG50 (i.e. TNG50-1), demonstrating good resolution convergence above the resolution limit threshold of each simulation. The top-right panel shows an inset corresponding to the relation between the major-to-minor axis ratio ($q$) and cloud size. In the bottom-left panel, we show the cloud mass-size relation, with the background colored by the average number of gas cells per gas cloud, and the bottom right panel shows the cloud distance-size relation, with the colored instead showing the total number of clouds in each given pixel. The most abundant clouds have masses of $\sim 10^{6}\, \rm{M_\odot}$ and sizes of $\sim 10$\,kpc. Lower mass clouds, with smaller sizes, tend to reside at smaller galactocentric distances.}
\label{fig:cloudBasicProp1}
\end{figure*}

This observed value is more than an order of magnitude above our most populous halo. However, several caveats exist in this comparison: the observations only report high-velocity clouds, i.e. only those clouds which satisfy a given velocity threshold, while we do not here apply any such cut in TNG50. Most clearly, the number of clouds in simulations depends on numerical resolution, and we expect more clouds at better resolution \citep[][Figure~\ref{fig:cloudBasicProp1}, and \textcolor{blue}{Ramesh \& Nelson in prep}]{nelson2020}. The number of identified clouds is greater if we relax the minimum number of cells threshold. In this case, the gray curve shows the median relation for the case where all clouds with at least one member Voronoi cell are considered, i.e. if poorly resolved clouds are also included. As seen, the sample size increases by a factor of $\sim 10$, with a largely similar trend with respect to stellar mass. We undertake a more realistic comparison with observations in Section~\ref{sec:comp_obs}.

\subsection{Physical properties of clouds in TNG50 MW-like galaxies}\label{sec:clouds_physical}

Figure~\ref{fig:cloudBasicProp1} quantifies the physical properties of our TNG50 cold clouds. In the top left panel, we show the probability distribution function (PDF) of cold cloud masses. The distributions for individual MW-like galaxies are shown with thin curves, colored by the mass of the central SMBH. The thick black curve shows the median of the entire galaxy sample, while the purple and yellow curves show the median of the lowest and highest octiles, dividing the sample into eight percentiles based on SMBH mass. 

The cloud mass distribution peaks around $10^{6.1}\, \rm{M_\odot}$, which is slightly more than ten times the average mass of baryon resolution elements (gas cells) in TNG50. The PDF drops sharply towards the left of this peak, as a result of our cloud definition and the resolution limit of the simulation. More massive cold clouds are rarer. No strong trend with respect to SMBH mass is seen in case of distributions of cloud masses. For reference, in gray, we also show the median distribution for the case where clouds down to one member cell are considered, offseting this distribution vertically downwards better visibility. The shape of the gray curve largely resembles the one in black, however with the peak shifted towards lower masses.

While masses of HVCs in the Milky Way halo are poorly constrained owing to a lack of distance estimates, recent studies suggest that the typical HI mass of clouds varies from $\lesssim 10^5 \rm{M_\odot}$ \citep{wakker2001, adams2013} to $\sim 10^7 \rm{M_\odot}$ \citep{thom2008}. With TNG50, we can thus start to study clouds for a similar mass regime as in the real Milky Way, with the caution of limited resolution towards the low mass end.

In the inset, we assess numerical resolution convergence. To do so, we compare the cloud mass distribution from TNG50-1 (i.e. TNG50, black) with its lower resolution counterparts, TNG50-2 (red; 8 times lower mass resolution) and TNG50-3 (orange; 64 times lower mass resolution). The curves correspond to the median across the sample of MW-like galaxies, matched between pairs of runs. Importantly, we see good convergence between the different runs above the resolution limit: namely, the number density of `massive' clouds is roughly independent of resolution. In each case, the mass function peaks, turns over, and then drops rapidly at roughly ten times the resolution limit of the simulation, corresponding to a minimum of ten member gas cells per cloud. As expected, this peak shifts towards higher masses for the lower resolution runs. This is directly analogous to the halo mass function in any cosmological simulation, where resolution convergence demands that the space number density (and properties) of haloes \textit{above the resolution limit} are in good agreement \citep[e.g][]{boylan2009,prada2012,springel2021}. Clouds (or haloes) can only exist above a given mass resolution threshold, and smaller structures are simply absent at lower resolution.

The top-right panel of Figure~\ref{fig:cloudBasicProp1} shows the cloud size distribution. To estimate sizes, we fit the vertices of Voronoi cells of clouds to an ellipsoid, and consider the size to be the geometric mean of the the lengths of the three axes. As before, we construct PDFs for each galaxy, all of which are shown with thin curves, colored by SMBH mass. The thick black curve corresponds to the median across the sample, while the purple and yellow curves are the median of the two extreme octiles in SMBH mass.

The cloud size distributions peak around $\sim 10$kpc, and the median PDF drops monotonically on either side. A weak trend of cloud sizes is visible as a function of SMBH mass: galaxies with less massive SMBHs have a slightly greater fraction of clouds in the low-size regime, and a slightly lower fraction of clouds in the high-size regime, as compared to galaxies with more massive SMBHs at their centres. We believe that this is linked to the radial distance-distributions of clouds in these haloes, which we return to in Section~\ref{sec:clouds_spatial_distr}. When the sample is split based on other quantities such as stellar mass or halo mass, a much weaker trend is observed. In gray, we also show the median distribution for the case where clouds with less than 10 member Voronoi cells are included, with the distribution vertically offset for better visibility. The shape of the gray curve largely resembles the one in black, however with the peak shifted towards smaller sizes.

In the inset, we show the relation between the major-to-minor axis ratio ($q$) and cloud size (x-axis), for the entire sample of clouds across all galaxies. Each point corresponds to an individual cloud, with color scaled in accordance to the mass of the cloud: the darkest points show clouds of mass $\sim 10^6 \rm{M_\odot}$, while the lightest correspond to $\sim 10^7 \rm{M_\odot}$. The gray curve shows the median. A clear trend is apparent, wherein small cold clouds are more `spherical', having lower values of $q$, in comparison to their more massive counterparts. The spherical nature of small clouds could be a result of poor resolution and/or be linked to the fact that \textsc{arepo} forces gas cells to be `round' \citep{springel2010}. A comparison with higher resolution simulations is required to truly assess the shapes of small clouds. Observationally, data suggests that HVCs typically have sizes of $\lesssim 10$kpc (e.g. \citealt{thom2008}), although the lack of distance measurements makes physical size inference challenging.

The bottom-left panel of Figure~\ref{fig:cloudBasicProp1} shows the relation between cloud mass and cloud size. We again stack all CGM clouds across the full sample of TNG50 MW-like galaxies. The median trend is shown with the solid gray curve, and the 16$^{\rm{th}}$ and 84$^{\rm{th}}$ percentile regions with dashed lines. Background color encodes the average number of member Voronoi cells per cloud, and these pixels are clipped outside the 16$^{\rm{th}}$ and 84$^{\rm{th}}$ percentile regions for visual clarity. On average, CGM cold clouds with larger sizes are also more massive, and are comprised of more gas cells, i.e. they are better resolved.

The bottom-right panel of Figure~\ref{fig:cloudBasicProp1} shows the relation between the galactocentric distance of clouds (normalised by the virial radius) and their physical size. As before, we stack all clouds across the sample. The solid curve shows the median relation, and dashed curves correspond to the 16$^{\rm{th}}$ and 84$^{\rm{th}}$ percentile regions. Background color shows the number of clouds in each corresponding bin, and pixels are clipped outside the percentile regions. On average, clouds get larger in size (and hence in mass, on average) with increasing distance. Most clouds are present in the inner half of the halo, and have sizes of $\sim 10$ kpc, consistent with the top-right panel. We suspect that this could be a result of increased ram-pressure stripping at smaller galactocentric distances, as a result of which `big' clouds are fragmented into smaller objects, thereby giving rise to a larger number of clouds in the inner halo, each of which are less massive in comparison to more distant counterparts.

\begin{figure*}
\centering 
\includegraphics[width=6.8cm]{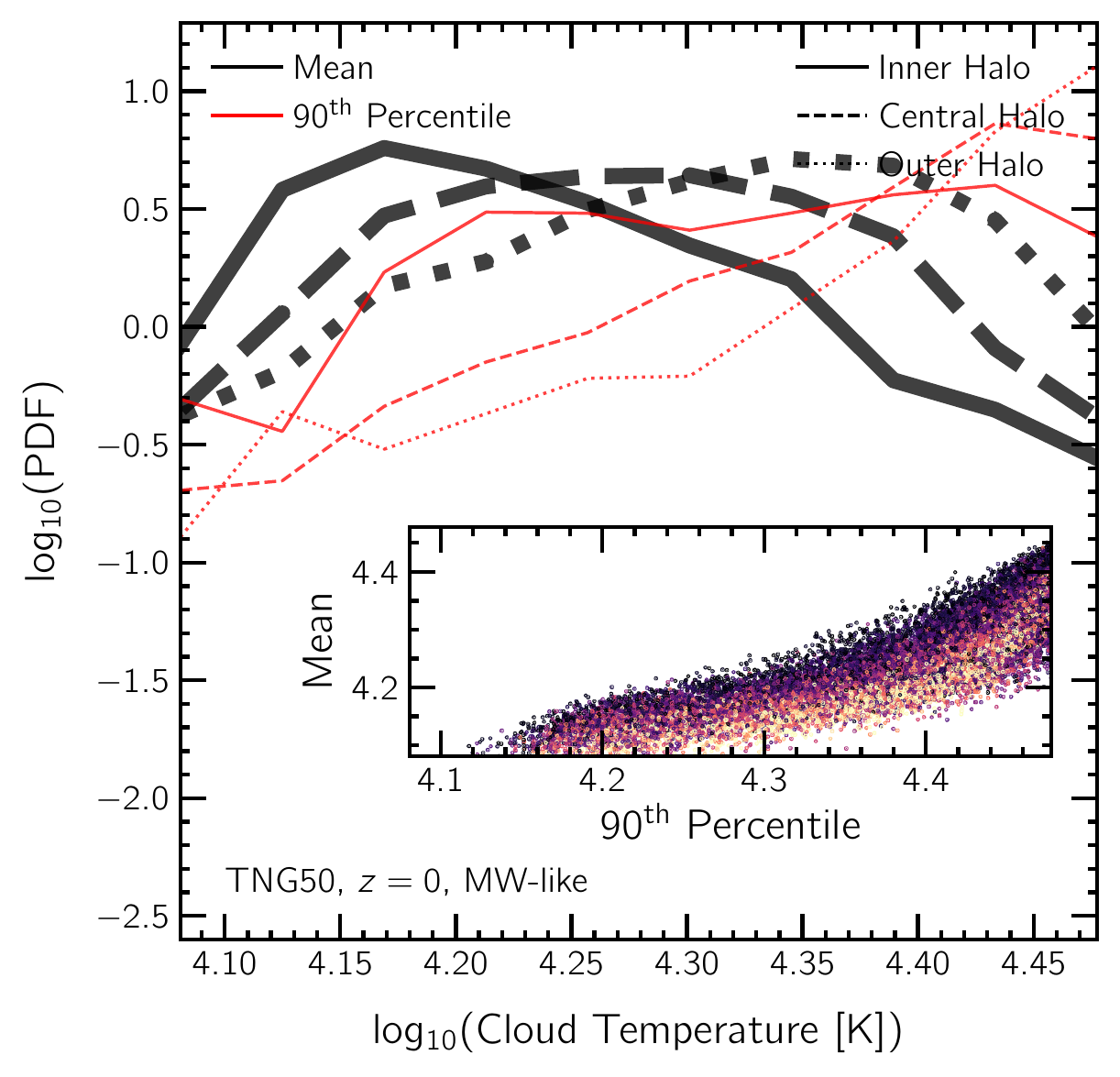}
\includegraphics[width=6.65cm]{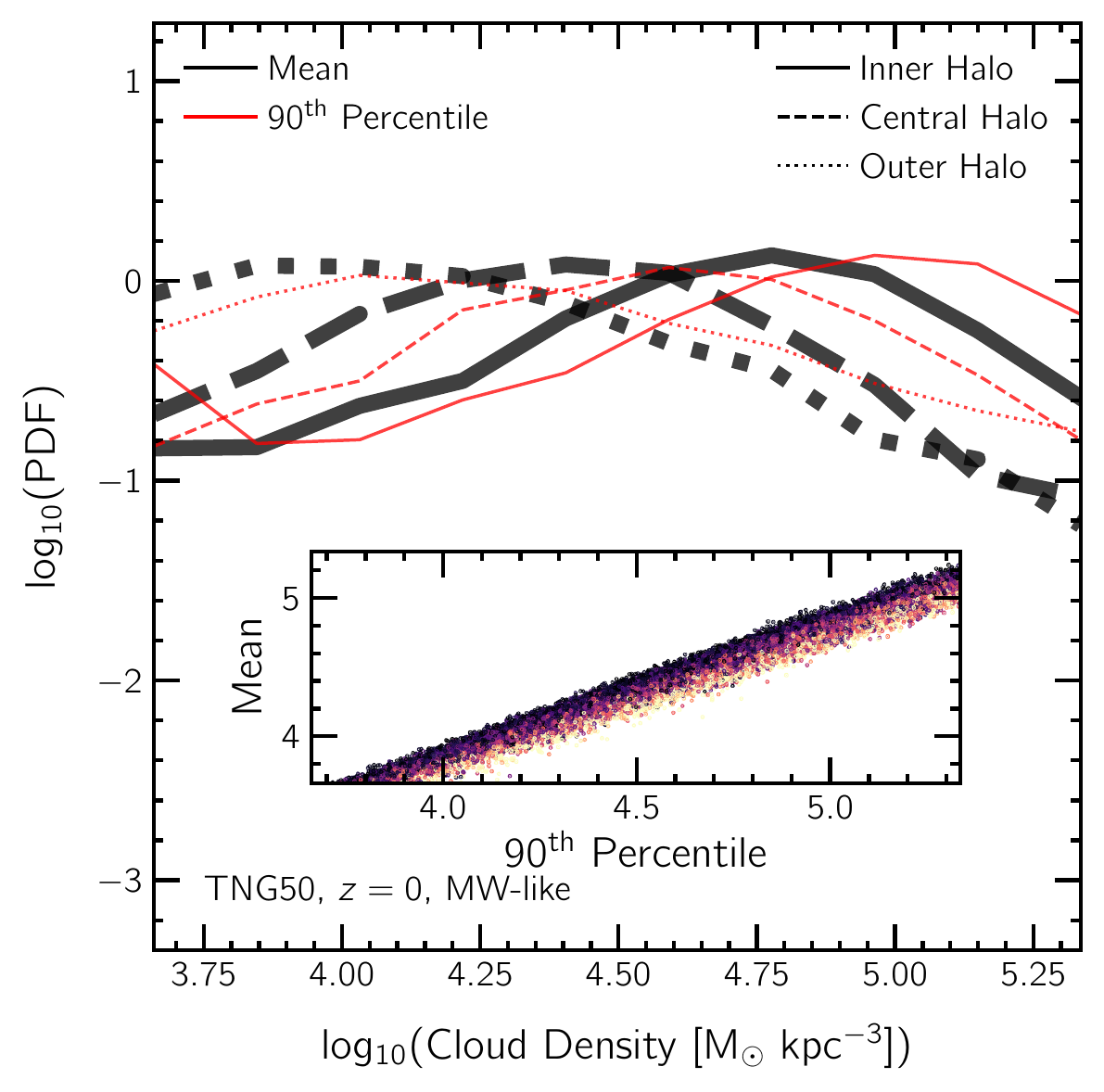}
\includegraphics[width=6.8cm]{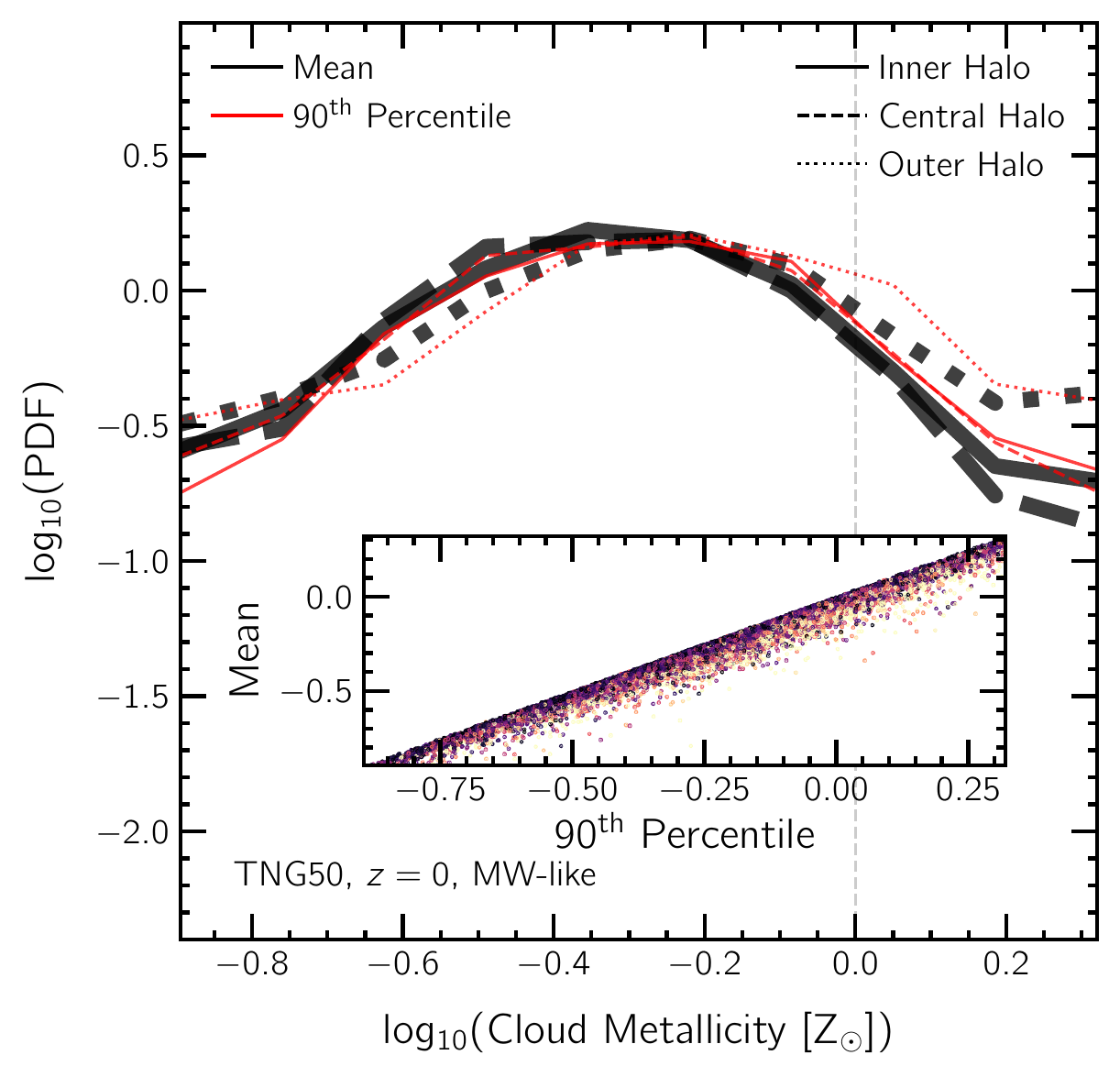}
\includegraphics[width=6.8cm]{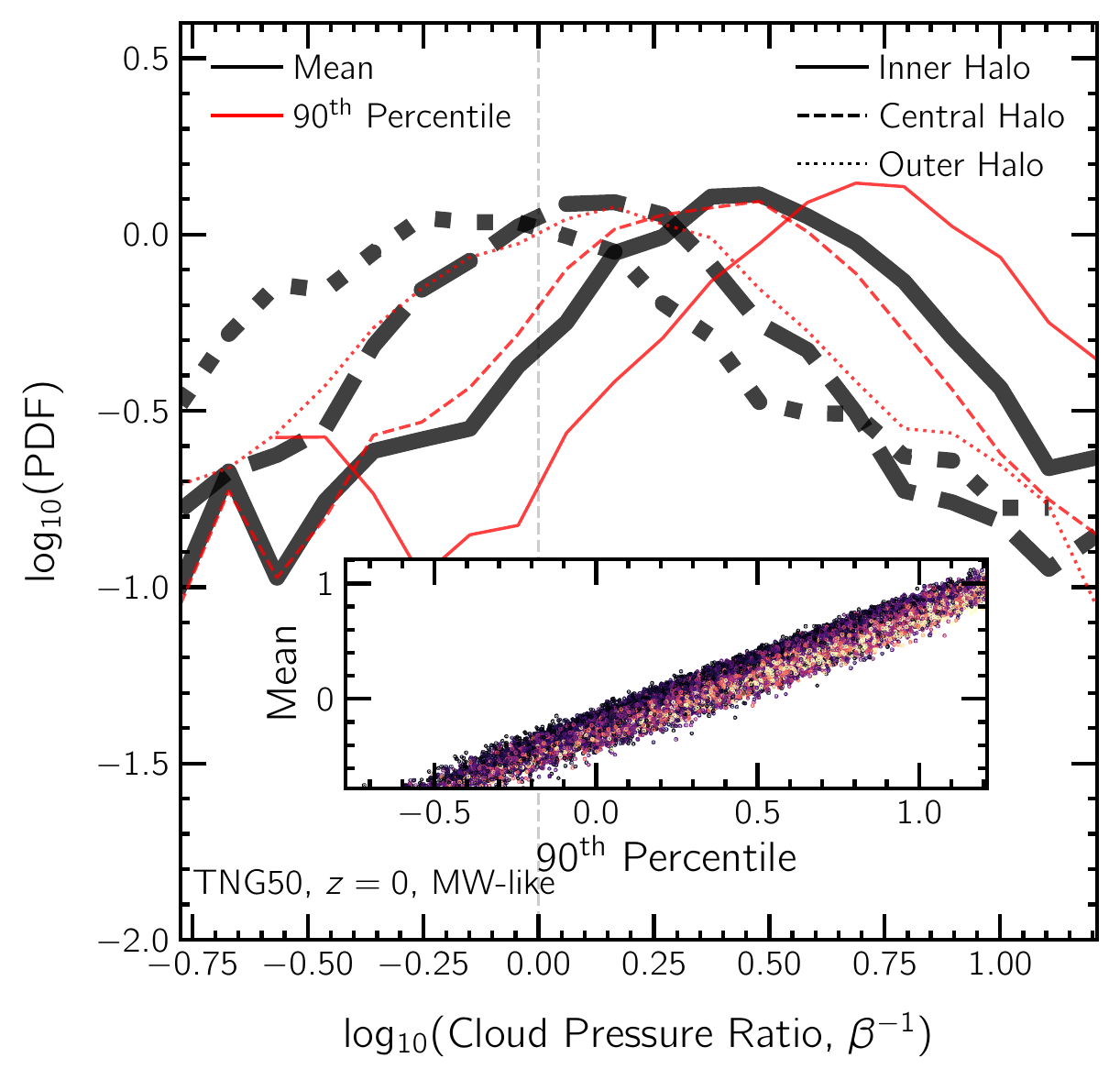}
\includegraphics[width=6.8cm]{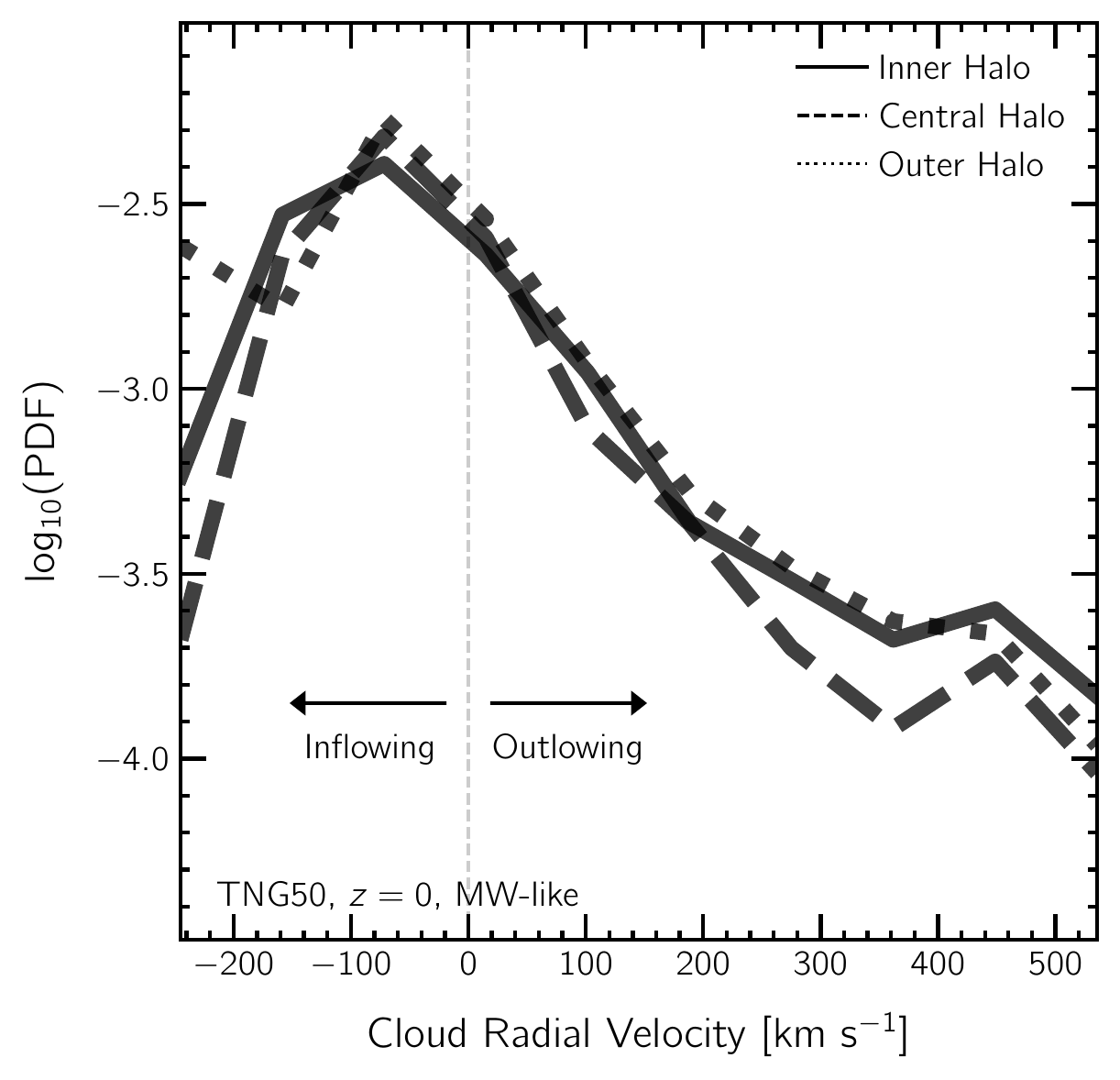}
\includegraphics[width=6.8cm]{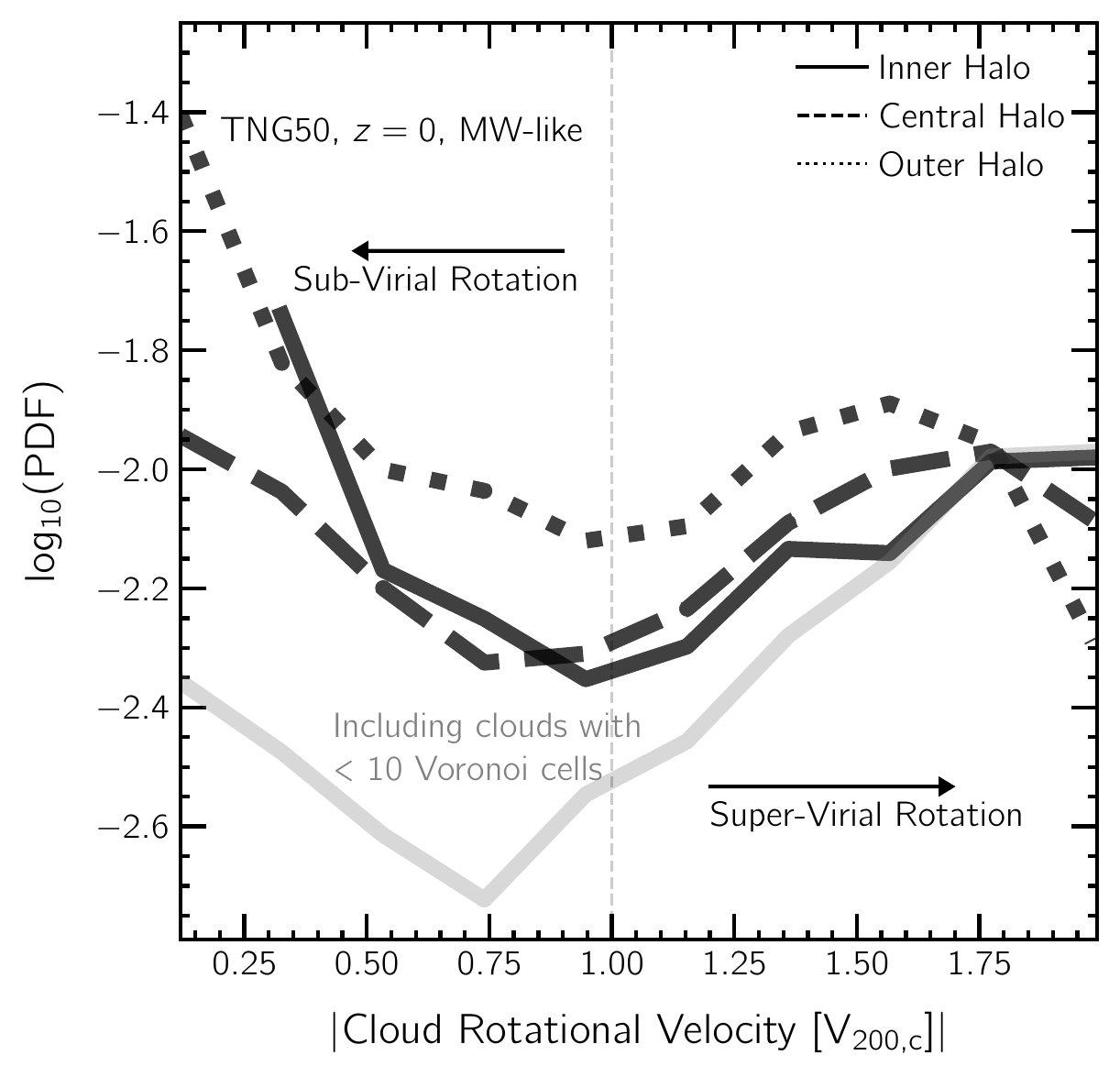}
\caption{Distributions of physical properties of clouds in the CGM of TNG50 MW-like galaxies. In each panel, we show the median behaviour across the full sample of galaxies, split into three regions of the halo: inner halo (solid curves; $0.15 < r/\rm{R_{200,c}} \leq 0.4$), central halo (dashed curves; $0.4 < r/\rm{R_{200,c}} \leq 0.7$) and outer halo (dotted curves; $0.7 < r/\rm{R_{200,c}} \leq 1.0$). The two upper panels show the distributions of temperature (left) and density (right): clouds closer to the centre of the halo are cooler and denser than those farther away. The centre-left and right-panels show the distribution of cloud metallicity and magnetic to thermal pressure ratio ($\beta^{-1}$), respectively: while clouds throughout the halo  mainly possess sub-solar metallicity and pressure ratios close to unity, clouds in the inner halo are the most likely to be highly enriched and dominated by magentic pressure. The bottom-left and -right panels show distributions of the radial velocity (in km/s) and rotational velocity (normalised by the virial velocity), respectively: a smaller fraction of clouds in the inner halo have dominant rotational motion and a greater fraction are either strongly inflowing or outflowing. Finally, the upper two rows also compare values derived in two ways: the mean value across the gas within each cloud (black; as in the lower panels), and the 90$^{\rm{th}}$ percentile value of the gas properties within each cloud (red). The insets show the relation between these two values, demonstrating that TNG50 clouds exhibit inner inhomogeneities in their metal content and temperature-, density- and pressure-structures.}
\label{fig:cloudBasicProp2}
\end{figure*}

\begin{figure*}
\centering 
\includegraphics[width=8.25cm]{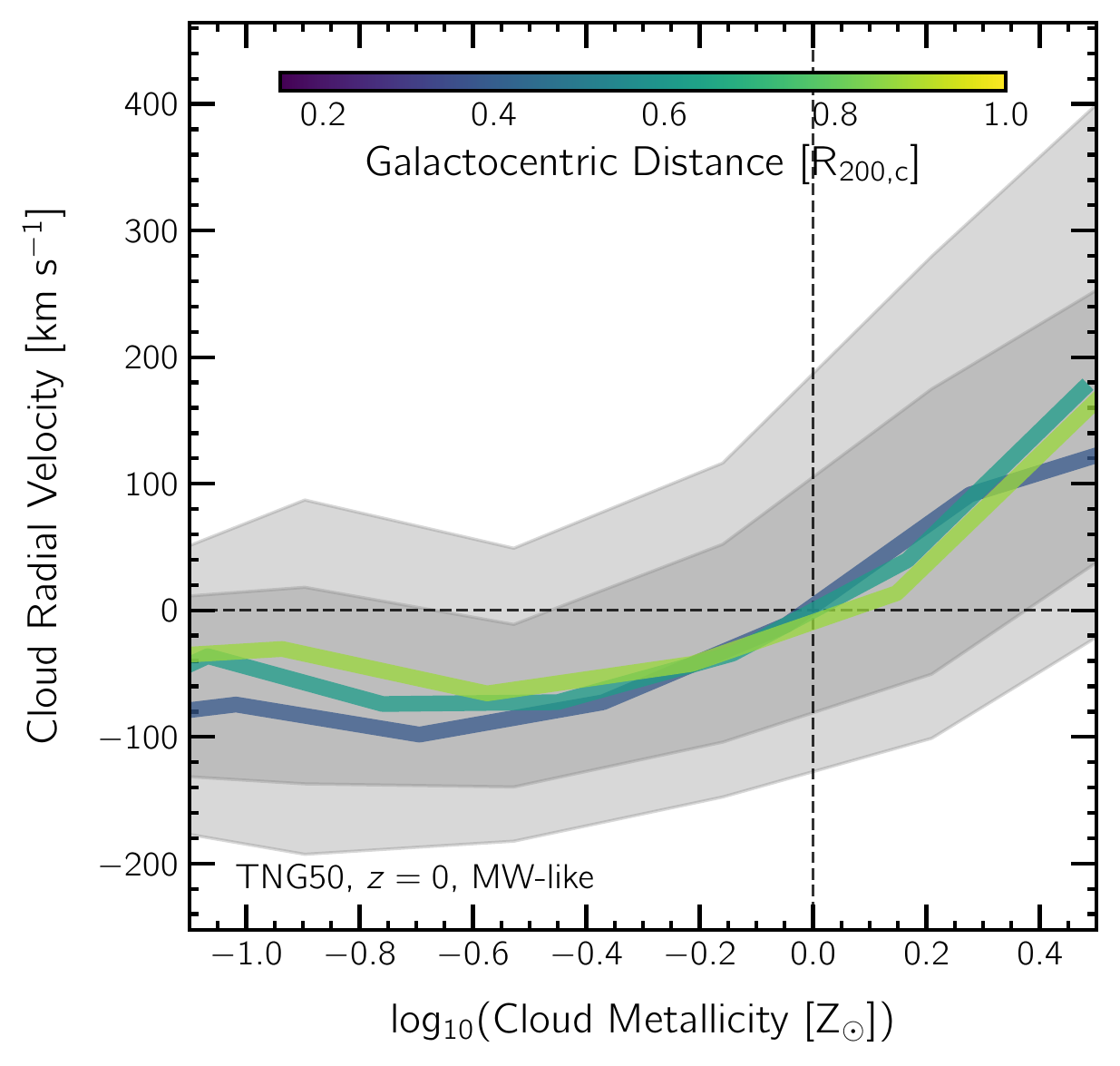}
\includegraphics[width=8cm]{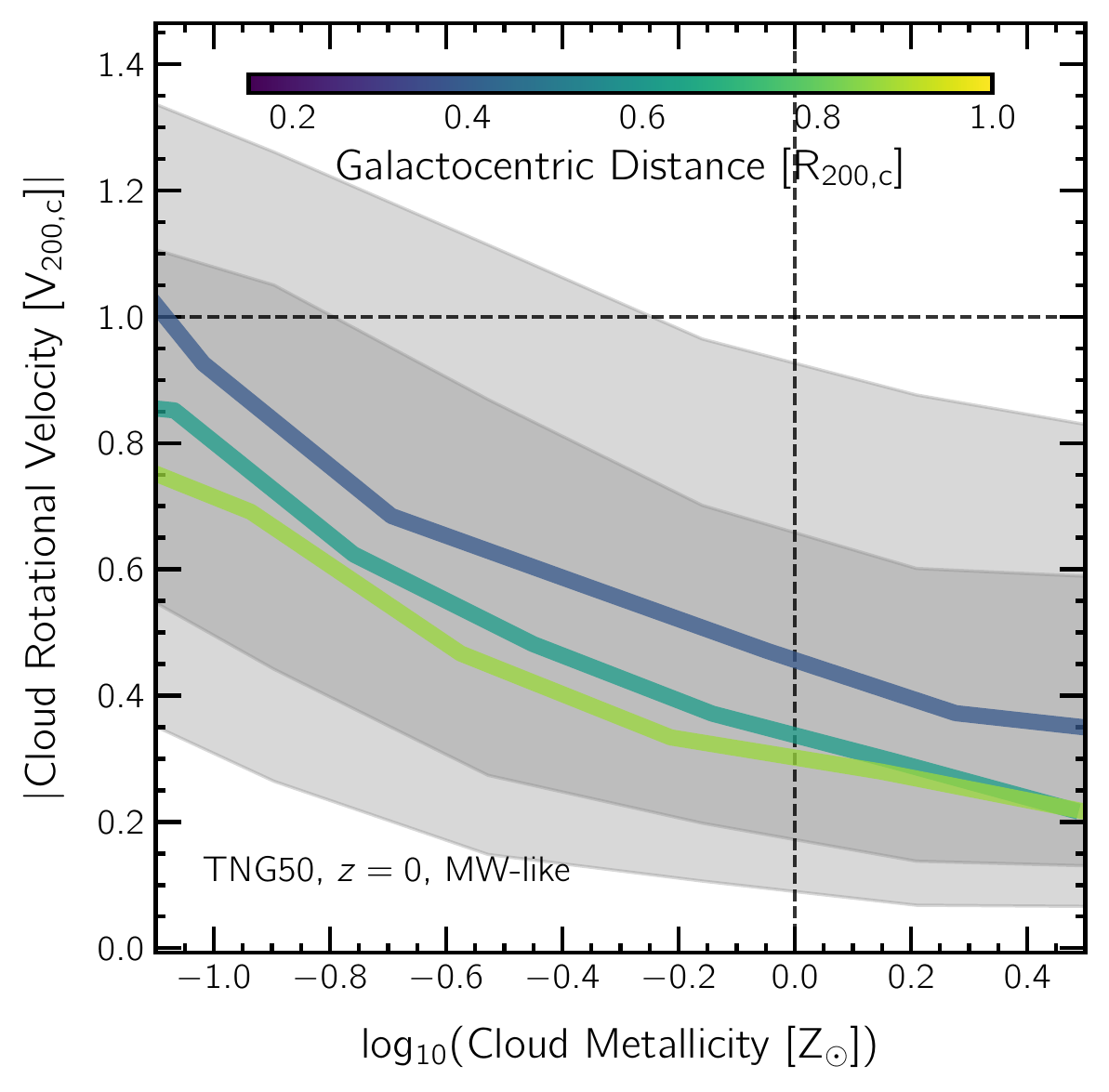}
\caption{Relation between kinematics of clouds and metallicity in the CGM of TNG50 MW-like galaxies. In the left (right) panel, we show the radial (rotational) velocity of clouds as a function of their metallicity. In both panels, the solid curves represent median values for three different mass bins, while the shaded regions correspond to percentile regions of the entire sample. On average, clouds with sub-solar metallicity are seen to be inflowing, with more pristine clouds inflowing at larger velocities, while clouds with super-clouds metallicity are preferentially outflowing.}
\label{fig:cloudKinProp}
\end{figure*}

In Figure~\ref{fig:cloudBasicProp2}, we consider several additional thermodynamical and physical properties of cold clouds in the CGM of TNG50 MW analogs: temperature, density, metallicity, pressure and radial and rotational velocities. As before, we construct individual PDFs for each galaxy, but show here only the median of these PDFs, for clarity. We further split each PDF into three components based on the galactocentric distance of the clouds: the inner halo (solid curves; $0.15 < r/\rm{R_{200,c}} \leq 0.4$), central halo (dashed curves; $0.4 < r/\rm{R_{200,c}} \leq 0.7$) and outer halo (dotted curves; $0.7 < r/\rm{R_{200,c}} \leq 1.0$). In the upper two rows, we consider two values for each cloud: the mean, and the 90$^{\rm{th}}$ percentile of the property, computed from all gas that comprises the cloud. The distributions of the former are shown in black curves, while the latter are shown in red. For these upper four panels, in insets, we also show the relation between the mean and 90$^{\rm{th}}$ percentile values for the stacked set of cold clouds across the entire sample. Each point corresponds to an individual cloud, with color scaled in accordance to the mass of the cloud: the darkest points correspond to clouds of mass $\sim 10^6 \rm{M_\odot}$, while the lightest correspond to $\sim 10^7 \rm{M_\odot}$.

In the top left panel of Figure~\ref{fig:cloudBasicProp2}, we show distributions of CGM cloud temperatures. As by construction we aim to identify {\it cold} gas structures ($T<10^{4.5}$K; see Section~\ref{cloud_algo}), we expect the cloud-wide average temperatures to be below this threshold value. Clouds in the inner halo have the `coldest' mean temperatures, with the distribution peaking at $\sim$\,$10^{4.15}$\,K. Clouds farther away from the centre are progressively more warm, with distributions shifting horizontally by $\sim$\,$0.1$ and $0.2$ dex for the central- and outer-halo, respectively. For each of the three regions of the halo, the distributions corresponding to the 90$^{\rm{th}}$ percentile values are skewed towards warmer temperatures, suggesting that clouds likely posses significant inhomogeneities in their inner temperature structure, consistent with \cite{nelson2020}. This behaviour is also well captured in the inset, which shows a large spread in the mean-90$^{\rm{th}}$ percentile plane.

In the top-right panel, we show distributions of the mass densities of the cold clouds. Clouds in the inner regions of the halo are most dense, with a peak in the density distribution at $\sim 10^{4.75}~\rm{M_\odot~kpc^{-3}}$. Clouds farther away from the centre are progressively more rarified, with peaks shifted by roughly $-0.4$ and $-0.8$ dex for the central and outer halo distributions, respectively. In all three regions of the halo, distributions of mean and 90$^{\rm{th}}$ percentiles are largely similar in shape, albeit the latter is skewed towards higher density values. Similar to the previous panel, the inset points towards an inhomogeneity in the inner density structure of clouds, albeit not as pronounced as their temperature structures.

In the centre left panel, we show distributions of cloud metallicity. The dashed vertical line placed at a x-axis value of 0 demarcates super-solar clouds from their sub-solar counterparts. Throughout the MW-like haloes, clouds with (mean) super-solar metallicity (i.e. clouds with $Z \geq Z_\odot$) are sub-dominant, and only account for $\sim 10.5$ per cent of the population of all clouds.  In the inner halo, the median cloud metallicity is $\sim 10^{-0.3}~Z_\odot$. In the central and outer regions of the halo, metallicity PDFs peak at marginally higher values, but are equally broad as the inner halo, stretching $\gtrsim 1$\,dex between the two extremes. The median PDFs show little difference between the mean and 90$^{\rm{th}}$ percentiles for cloud metallicities, although a deviation between the two values is seen in the inset for massive clouds, i.e. well resolved clouds are typically not homogeneous with respect to their metallicity content. 

On the observational end, a large range of cloud metallicities have been inferred. For instance, \cite{richter2001} estimate the metallicities of Complex C and IV Arch to be $\sim 0.1~Z_\odot$ and $\sim 1~Z_\odot$ respectively, while \cite{collins2003} and \cite{tripp2003} propose a slightly higher upper limit of $\sim 0.3 - 0.6~Z_\odot$ for Complex C. \cite{zech2008} and \cite{yao2011} report the possible existence of highly super-solar clouds with metallicities of $\sim 1.65~Z_\odot$ and $\sim 2.08~Z_\odot$ respectively, although the uncertainties on these measurements are quite large. On the other end of the metallicity spectrum, \cite{tripp2012} report that the metallicity of HVC gas in the direction of the gaseous `outer arm' is typically sub-solar, possibly around $\sim 0.2 - 0.5~Z_\odot$. The metal content predicted by TNG50 for the cold clouds around MW-like galaxies encompasses all these observationally-inferred metallicity values.

The centre right panel of Figure~\ref{fig:cloudBasicProp2} shows distributions of the magnetic to thermal pressure ratio, $\beta^{-1}$, of clouds. Overall, $\sim65.7$ per cent of cold clouds in TNG50 MW-like haloes are dominated by magnetic rather than thermal pressure (i.e. $\beta^{-1} > 1$). This is in marked contrast to the case of more massive haloes (M$_{\rm{200,c}} \gtrsim 10^{13} \rm{M_\odot}$) in the same TNG50 simulation, where magnetic pressure is clearly more dominant with respect to its thermal component for almost the entire population of clouds \citep{nelson2020}. We suspect that this is a direct effect of the reduced magnetic field strengths in the haloes of MW-like galaxies \citep[see also][]{faerman2023}, as compared to their more massive counterparts \citep{marinacci2018}. The PDF for the inner halo around TNG50 MW-like galaxies has a maxima at $\beta^{-1} \sim 2.8$, and is skewed towards positive values of log$_{10}$($\beta^{-1}$). However, magnetic pressure does not dominate to this extent for clouds in the central and outer regions of the halo. This is likely a consequence of the radial dependence of magnetic field strengths in MW-like haloes \citep{ramesh2022}. Finally, unlike the case of metallicity, a difference between the distributions of the mean and 90$^{\rm{th}}$ percentile values is visible: a horizontal offset by $\sim 0.3$ dex is present in all three regions of the halo. The inset shows that a larger spread in the mean-90$^{\rm{th}}$ percentile plane is present, suggesting that some regions of clouds are somewhat dominated more by magnetic pressure than others, similar to what was seen in Figure~\ref{fig:cloudAlgo2}.

Finally, in the lower panels of Figure~\ref{fig:cloudBasicProp2}, we quantify how the cold clouds move through the CGM of MW-like galaxies, in terms of their radial velocity, i.e. the component of the velocity vector that is oriented parallel to the position vector, is shown on the left, and rotational velocity, i.e. the component of the velocity vector along the plane orthogonal to the position vector, on the right.

Across the whole cold cloud sample, TNG50 predicts a relatively robust mix of outflowing versus inflowing cold clouds. Overall, outflowing clouds are sub-dominant, accounting for $\sim27.5$ per cent of the population. A dependence of radial velocity distributions on distance is present: although all three PDFs peak around the same value, the tails portray different behaviours at the high-velocity end. The inner halo has a slightly larger fraction of clouds moving radially at larger velocities, both in the inflowing and outflowing directions. The central and outer haloes have similar fractions of clouds at the high inflow velocity end. However, the central halo hosts a relatively larger fraction of cold clouds outflowing at large velocities, in comparison to the outer halo.

Direct comparison with kinematic results from observations is difficult, since only the velocity component along the line of sight to the observer is accessible, and not the radial velocity with respect to the centre of the Milky Way. However, \cite{moss2013} do provide velocities transformed into the frame of reference of the galactic center, and we note that $\sim 41$ per cent of clouds in their catalog are outflowing with respect to the center of the Milky Way.

The rotational velocities of the clouds in the lower right panel of Figure~\ref{fig:cloudBasicProp2} are normalised by the virial velocity of the corresponding halo. In the inner halo, according to TNG50, a smaller fraction of cold clouds are rotating at velocities in excess of the virial velocity, as compared to the other two regions of the halo. This is consistent with the previous panel that shows that a larger fraction of clouds in the inner halo have higher radial velocities: their motion is likely dominated by feedback mechanisms, and less by gravity. Cold clouds at larger distances are more likely to be rotating at super-virial velocities, as compared to the inner halo. In all regions of the halo, clouds with sub-virial rotational velocities are dominant. However, no clear monotonic trend with respect to distance is present at small values of rotational velocities.

While all panels except the bottom-right remain qualitatively unaffected if the threshold on minimum number of cells per cloud is reduced (not shown explicitly), considering clouds with less than ten member cells tilts the balance between sub-virial and super-virial rotation: to demonstrate this, in the solid gray curve, we show the median relation for clouds in the inner halo, had all clouds down to one Voronoi cell per cloud been considered. Clearly, the (relative) fraction of clouds with sub-virial rotation is lower in comparison to super-virial counterparts. Similar trends exist for the central and outer regions of the halo as well (not shown). Higher resolution simulations are required to determine if this behaviour is a consequence of limited resolution, or if clouds of lower masses (i.e $\lesssim 10^6$ M$_\odot$) have physically different kinematics and rotation.

\subsection{Relationship between kinematics and physical properties}

Kinematics and physical cloud properties are likely related. Observational studies have noted a correlation between metallicity of clouds and their radial velocities \citep[e.g.][]{richter2001}, such that metal-rich clouds are preferentially outflowing, and thus likely originate from the galaxy. On the other hand, metal-poor clouds are instead preferentially associated with inflows, and so are likely of non-galactic origin and possibly tracing relatively less enriched cosmological gas accretion \citep{nelson2013}.

In Figure~\ref{fig:cloudKinProp}, we explore these correlations for TNG50 clouds. In both panels, we stack all clouds across our sample of MW-like galaxies, and split them into three bins based on distance: inner halo ($0.15 < r/\rm{R_{200,c}} \leq 0.4$), central halo ($0.4 < r/\rm{R_{200,c}} \leq 0.7$) and outer halo ($0.7 < r/\rm{R_{200,c}} \leq 1.0$). Median values for each of these bins are shown through solid curves, and colored by the mean distance corresponding to each bin, while shaded regions show the 16$^{\rm{th}}$-84$^{\rm{th}}$ and 5$^{\rm{th}}$-95$^{\rm{th}}$ percentile regions for the entire sample. 
 
The left panel shows the radial velocities of clouds as a function of their (mean) metallicities. A monotonic trend is seen, wherein radial velocities increase with increasing metallicity. Interestingly, all three medians cross a value of $0$ km s$^{-1}$ as they reach super-solar metallicity. Medians corresponding to larger distances are more steep in the positive radial velocity regime, and are less steep in the negative radial velocity regime. In the right panel, we show the rotational velocity of clouds (normalised by the virial velocity) as a function of cloud metallicity. All three median curves monotonically shift to higher metallicites at lower rotational velocities. That is, cold clouds with more radial and less rotational motions are more metal enriched. TNG50 therefore suggests that highly-enriched cold clouds are dominated by outflows and feedback-driven motions, rather than having gravitationally-induced dynamics only, while the opposite holds for their more pristine counterparts.

\begin{figure*}
\centering 
\includegraphics[width=16.5cm]{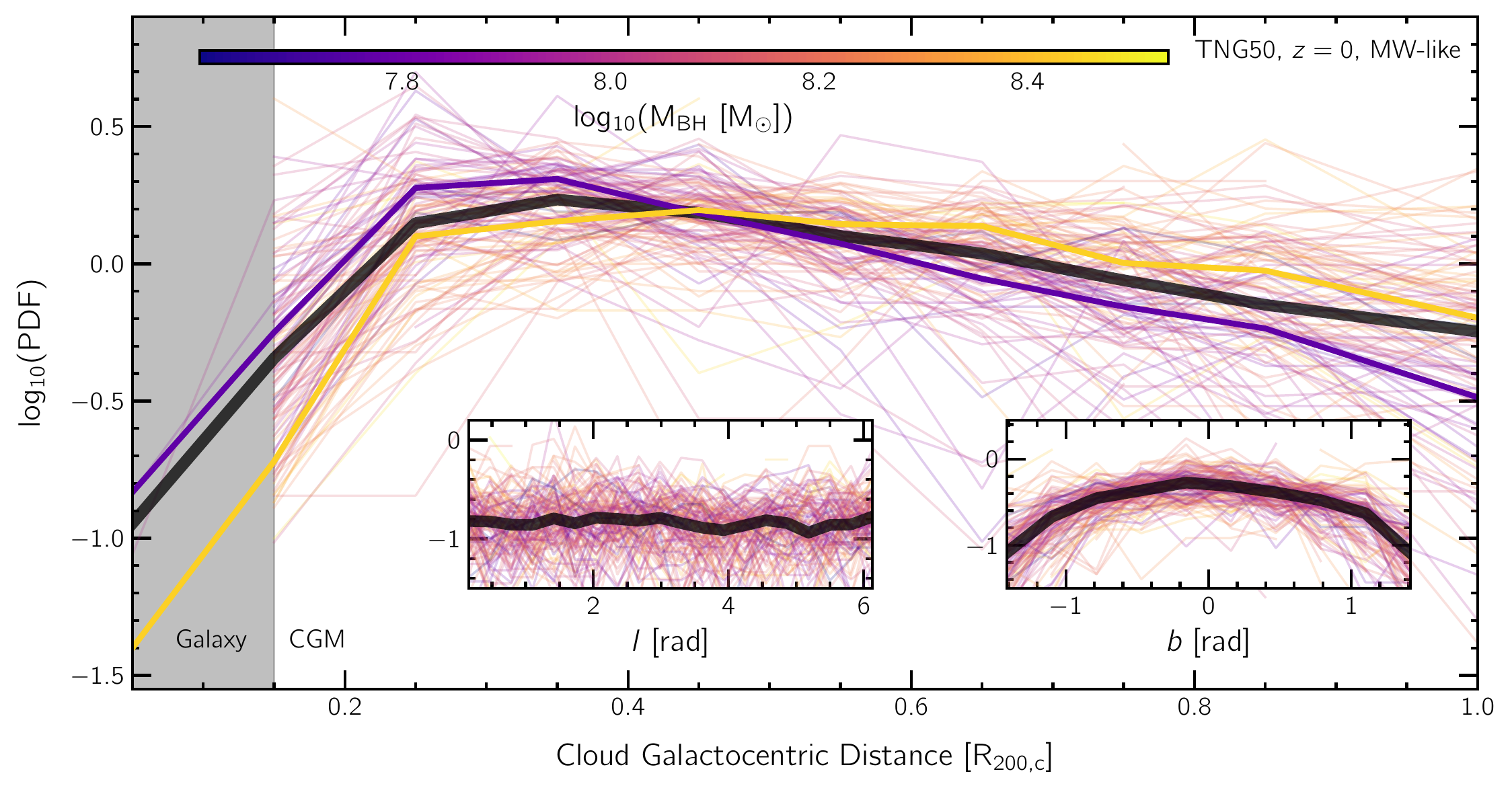}
\caption{The spatial distributions of clouds in the CGM of TNG50 MW-like galaxies. The main panel shows distributions of the galactocentric distance of clouds (normalised by the virial radius of the halo, R$_{\rm{200,c}}$). Cold clouds are found throughout the CGM, particularly at intermediate radii. Individual galaxies are shown with thin curves, colored by the mass of the central SMBH, while the median of the sample is shown in the thicker black curve. In the main panel, we also show medians of the two extreme octile regions. A trend with SMBH mass is present, wherein galaxies with less massive SMBHs have a greater number of clouds at smaller galactocentric distances. The left (right) inset shows the distribution of longitude (latitude) of the cloud population. While the medians in each case are more or less smooth, the individual curves are noisy, suggesting that clouds are clustered in specific locations.}
\label{fig:cloudSpatDistr}
\end{figure*}

\subsection{Spatial distribution of clouds throughout MW-like haloes}\label{sec:clouds_spatial_distr}

Having quantified the basic properties of CGM cold clouds in TNG50 MW-like galaxies, we now turn to the distribution of clouds throughout the haloes, and how their properties with respect to their local surroundings depend on galactocentric distance. 

In Figure~\ref{fig:cloudSpatDistr}, we show the distributions of the positions of cold clouds: the main panel focuses on the three-dimensional radial distance, while the two insets show distributions of longitude ($l$) and latitude ($b$). PDFs of individual MW-like haloes are shown as thin curves, colored by the mass of the SMBH at the centre of each galaxy. The median of all these PDFs is shown with the thick black curve. In the main panel, we also show medians or MW-like galaxies with over- and under-massive SMBHs (two extreme octiles, thick yellow and purple curves).

\begin{figure*}
\centering 
\includegraphics[width=7.3cm]{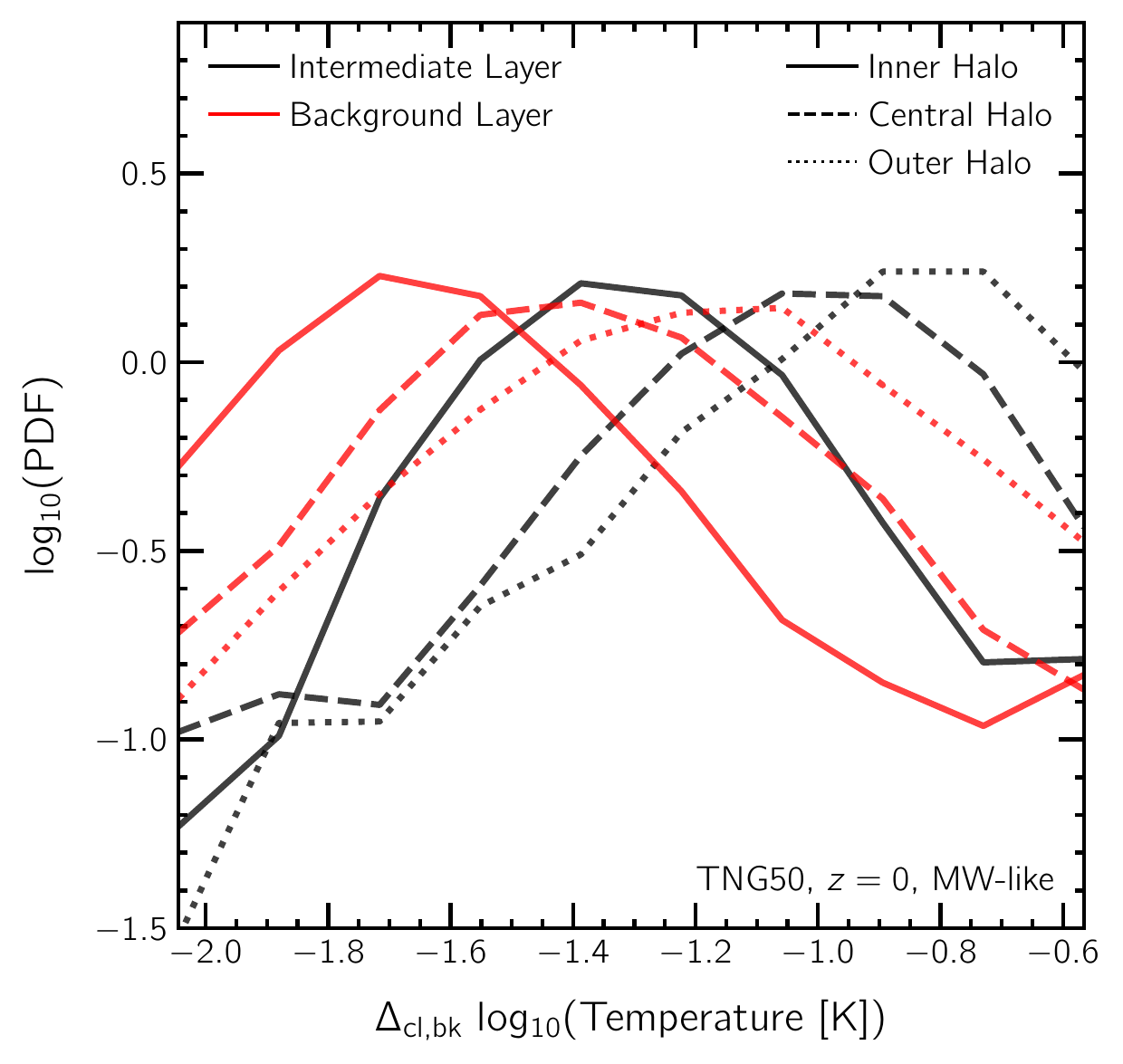}
\includegraphics[width=7.3cm]{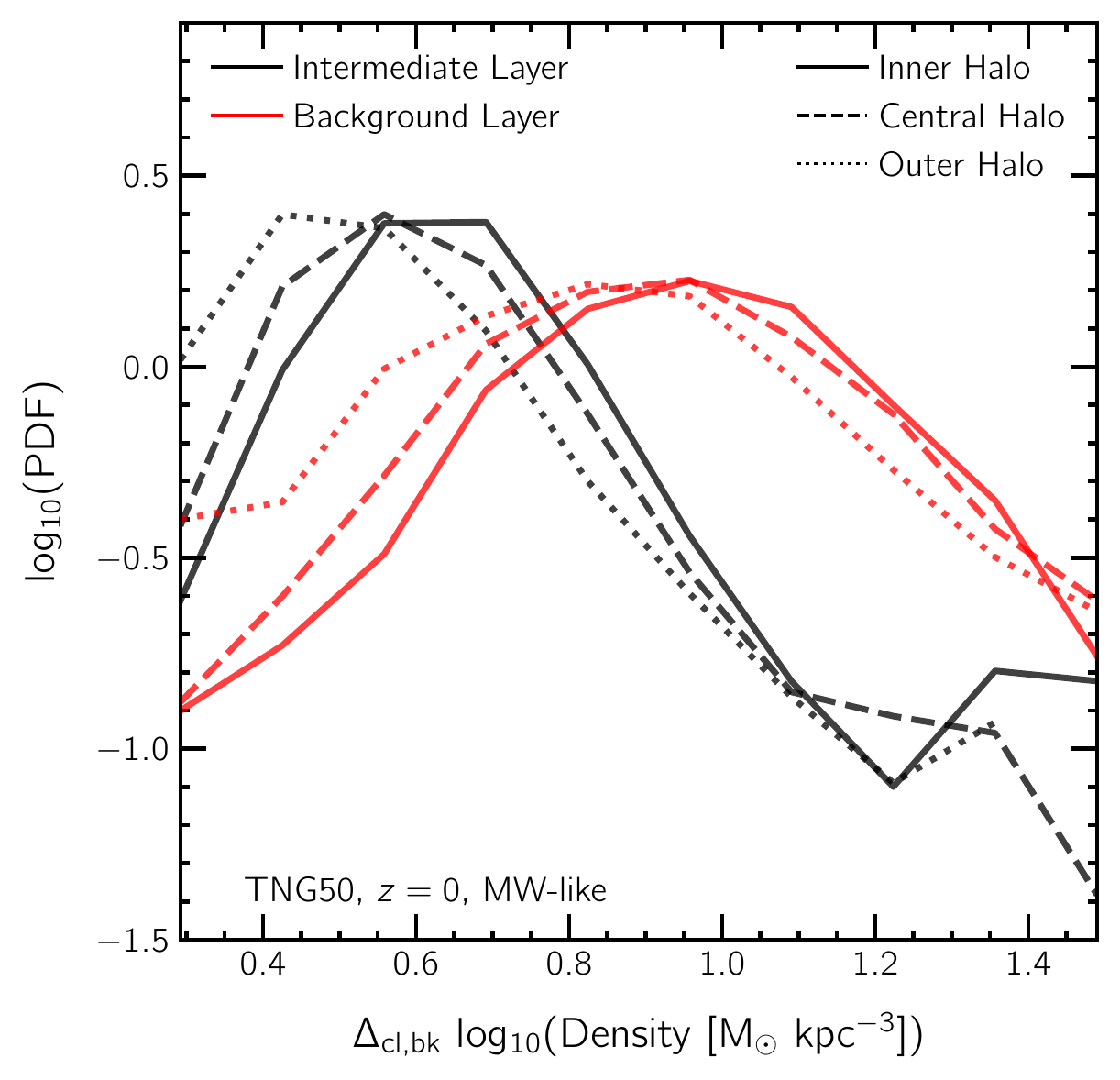}
\includegraphics[width=7.3cm]{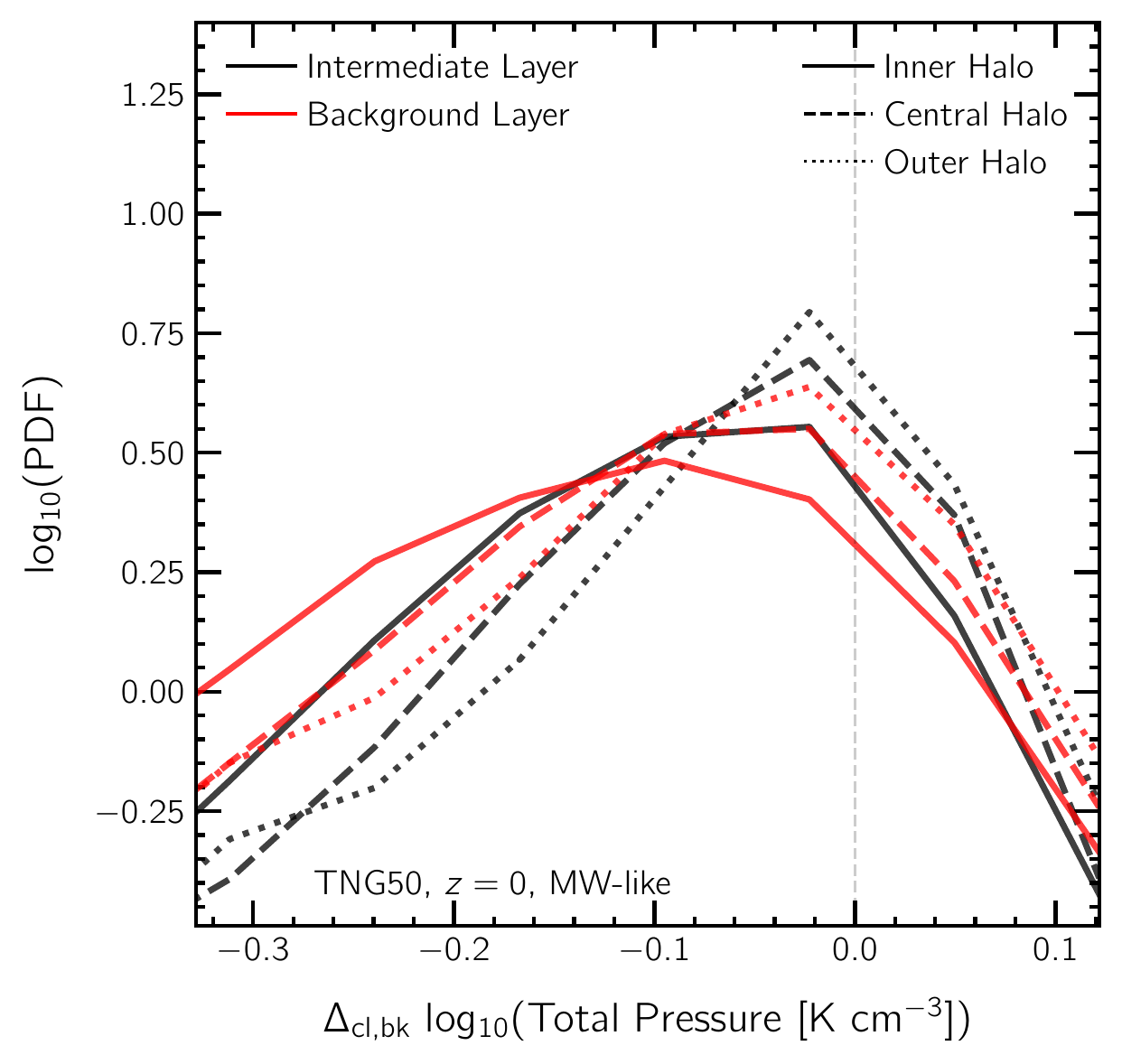}
\includegraphics[width=7.3cm]{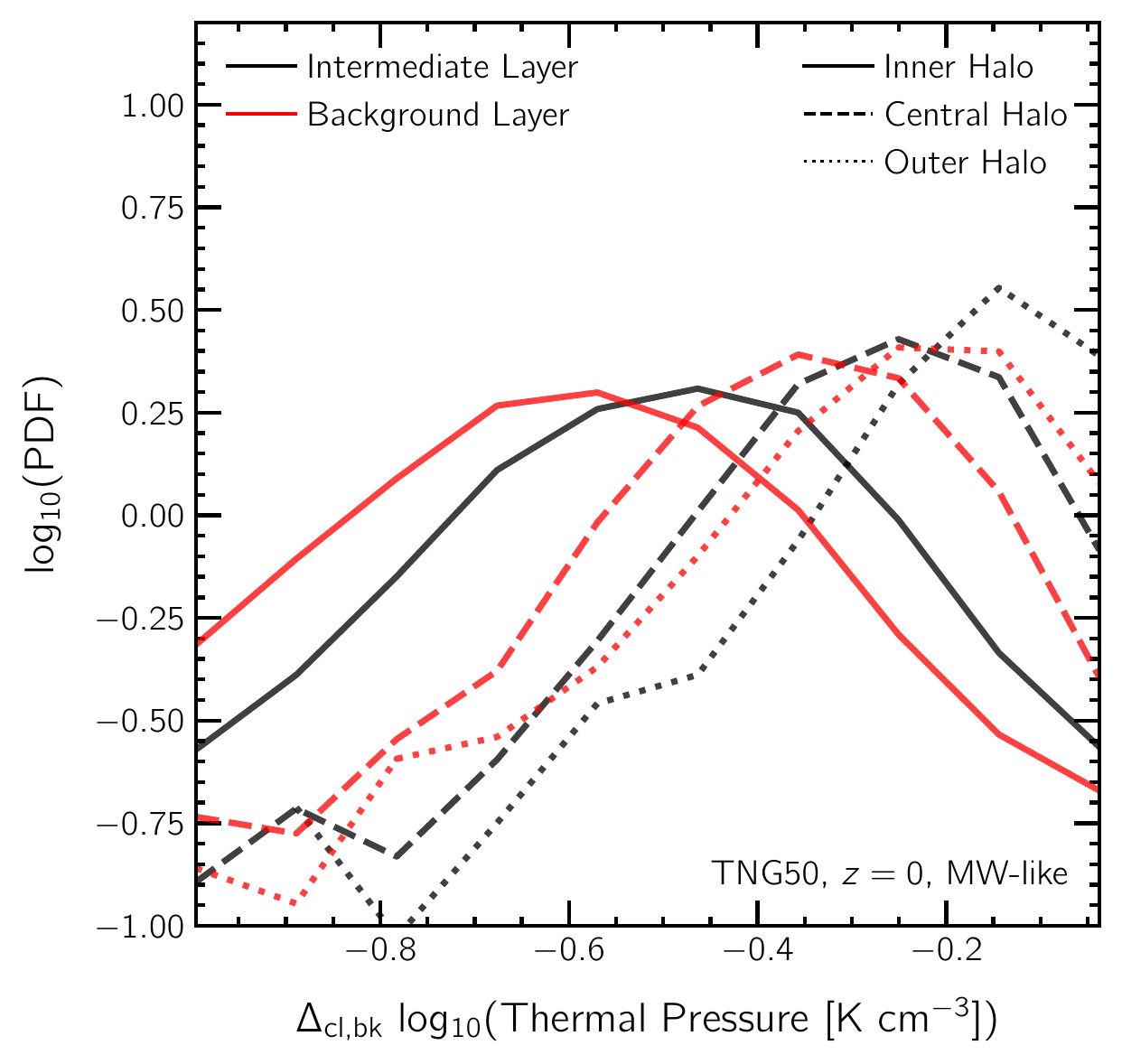}
\includegraphics[width=7.3cm]{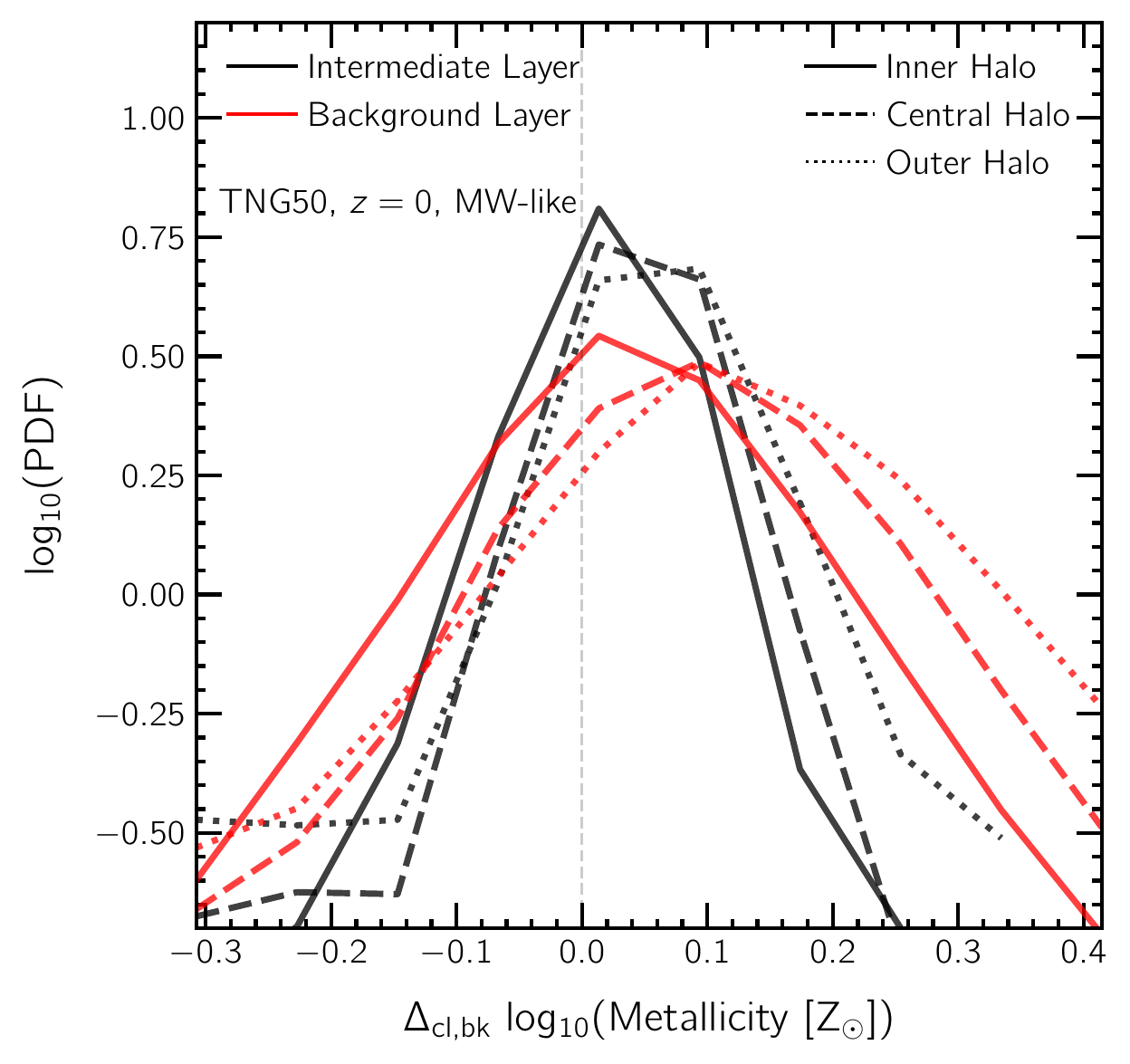}
\includegraphics[width=7.3cm]{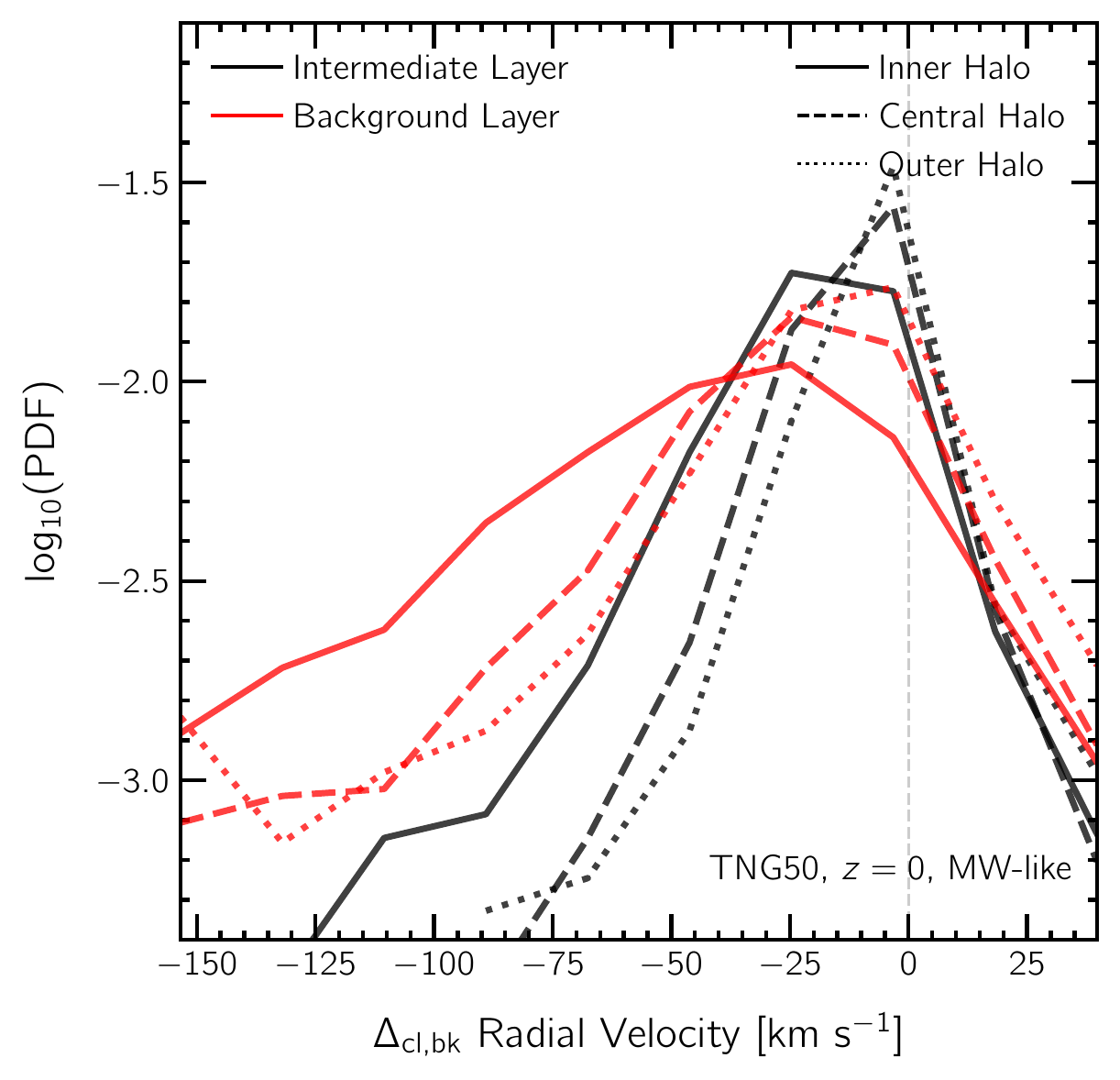}
\caption{Distributions of contrast in properties of TNG50 MW-like clouds with respect to their intermediate and background layers, i.e. with respect to the local, ambient CGM gas. In each panel, distributions of contrasts corresponding to the former are shown in black, while the latter are shown in red. We show median PDFs for three distance regimes: inner halo (solid curves; $0.15 < r/\rm{R_{200,c}} \leq 0.4$), central halo (dashed; $0.4 < r/\rm{R_{200,c}} \leq 0.7$) and outer halo (dotted; $0.7 < r/\rm{R_{200,c}} \leq 1.0$). Contrasts of temeprature are shown in the top-left panel, density in the top-right, total pressure in the centre-left, thermal pressure in the centre-right, metallicity in the bottom-left, and radial velocity in the bottom-right. In all cases, contrasts are more enhanced with respect to the background as compared to the intermediate layer, both in terms of the median, and the width of the distributions.}
\label{fig:cloudPropVsBackPropHist}
\end{figure*}

Cold clouds within the inner boundary of the CGM, i.e. $< 0.15 \times \RVIR$, are rare -- most cold gas mass in those regions makes up the galactic disk itself which is excluded by construction by our cold cloud finding algorithm (Section~\ref{cloud_algo}). On average across the galaxy sample, TNG50 clouds are most commonly found around $0.25 - 0.4 \times \RVIR$, and become progressively more rare at larger distances, as already noted for the galaxy of Figure~\ref{fig:cloudPosSatPos}.

Individual-galaxy curves are clearly `noisy', not as a result of low-number statistics, but rather because clouds cluster in specific locations. In addition, a strong trend is present with respect to SMBH mass: galaxies with less massive SMBHs at their centres have a larger fraction of their clouds at smaller distances, while the inverse is true for galaxies with more massive SMBHs. Similar to the trend with respect to sSFR in Figure~\ref{fig:numClouds}, we speculate this could be a result of kinetic winds of more massive SMBHs, which (i) drive (cold) gas away from the centre \citep[e.g.][]{zinger2020,ayromlou2022}, and/or (ii) destroy clouds or inhibut their formation at small galactocentric distances, either directly via hydrodynamical interactions, or indirectly by heating. HVCs within the real Milky Way are thought to have a similar distance trend, with clouds reducing in number towards outer regions of the halo \citep[][although this inference requires theoretical distance modeling]{olano2008}.

The left inset of Figure~\ref{fig:cloudSpatDistr} shows the distribution of cold clouds as a function of galactic longitude. As before, individual curves depicting individual galaxies show a large degree of diversity, primarily due to cold clouds being grouped around each other at specific longitudes. Such a feature is also seen in observations of the Milky Way halo: for instance, in the all-sky NHI map of \cite{westmeier2018}, clouds are not uniformly distributed in longitude, but rather are more concentrated at a subset of longitude ranges. 

The inset on the right similarly shows the distribution of clouds as a function of galactic latitude. We see a preference for a greater abundance of clouds at low latitudes, indicating alignment with the plane of the gaseous galactic disk. Specifically, $[50, 75, 90]$ per cent of all cold clouds are within latitudes of $~[30, 49, 66]$ degrees. A similar trend has also been observed in the Milky Way halo, with a strong concentration of HVCs close to the galactic plane \citep{moss2013}.  

\subsubsection{TNG50 cold clouds in relation to their ambient CGM}

In Figure~\ref{fig:cloudPropVsBackPropHist}, we show how cold cloud properties compare to their local background environment, and how they vary with distance. Since properties of gas, in general, vary with distance throughout the halo of TNG50 MW-like galaxies \citep{ramesh2022}, we focus on differences of properties of clouds with respect to the properties of the local surrounding: these are labelled as $\Delta_{\rm{cl,bk}}$, i.e. they are the (mean) value of a given property of a cloud minus the (mean) property of the surrounding. For the background, we consider two different definitions of layers of gas around each cloud\footnote{Each is identified geometrically using the connectivity of the Voronoi gas tessellation and the naturally connected Voronoi neighbors of cloud cells. Our background layers therefore have a one-cell thickness roughly equivalent to the numerical resolution, and reflect the irregular shapes of clouds.}: (i) the layer immediately surrounding the cloud, i.e. the `intermediate layer', shown throughout in black, and (ii) the next more outwards layer, which surrounds the intermediate layer, i.e. the true `background layer', shown in red. As in previous figures, in each panel, we construct PDFs for individual galaxies and show their median, but split into three different distance regimes: inner halo (solid curves; $0.15 < r/\rm{R_{200,c}} \leq 0.4$), central halo (dashed curves; $0.4 < r/\rm{R_{200,c}} \leq 0.7$) and outer halo (dotted curves; $0.7 < r/\rm{R_{200,c}} \leq 1.0$). 

In the upper-left panel of Figure~\ref{fig:cloudPropVsBackPropHist}, we show distributions of temperature contrasts between clouds and their surroundings. At all distances, clouds are cooler than both their intermediate and background layers, which is expected since our cloud-finding algorithm specifically selects cold gas to be part of clouds. A distance trend in values of temperature-contrasts is present: in the inner halo, on average, clouds are cooler than their intermediate layer by $\sim 1.4$\,dex, which reduces to $\sim 1.1$\,dex in the central halo, and further to $\sim 0.8$\,dex in the outer halo. Contrasts with respect to the background layer are more pronounced: in all three regions of the halo, these distributions are offset by roughly $-0.3$\,dex with respect to the distributions corresponding to the intermediate layer. This is consistent with earlier studies of TNG50 cold clouds \citep{nelson2020}, and with Figure~\ref{fig:cloudAlgo2}, wherein an intermediate mixing layer of warm gas is believed to be sandwiched between cold clouds and their hot backgrounds.

The upper-right panel shows a fundamental quantity of cold clouds in hot media: the density contrast, often written as $\chi = n_{\rm cold} / n_{\rm hot}$ \citep[e.g.][]{scannapieco2015}. At all distances, TNG50 cold clouds are denser than both their surrounding intermediate and background layers. A halocentric distance trend is present, wherein the over-density of clouds with respect to their surroundings is greater at smaller galactocentric distances: with respect to their intermediate layers, cold clouds are denser than the surroundings by $\sim 0.65$ dex in the inner halo and by $\sim 0.55$ and $0.45$ dex in the central and outer regions of the halo, respectively. Density contrasts with respect to the background layers are larger: $\sim 0.95$, $0.9$ and $0.85$ dex in the inner, central and outer halo, respectively. However, at all distances, distributions corresponding to contrasts with respect to the background are again broader, versus the intermediate layer. These findings are relevant in the context of previous, idealized numerical experiments. In fact, wind tunnel simulations generally assume high values of density contrasts, typically from $\sim 10$ to $\sim 1000$ \citep[e.g.][]{schneider2017}. Within the CGM of TNG50 MW-like galaxies, we find such large density contrasts rarely: they make up the tail of the distribution only. Our result suggests that typical cold clouds in MW-like haloes may have much weaker overdensities with respect to the background CGM, which would impact mixing efficiencies and overall survivability.

In the centre-left panel of Figure~\ref{fig:cloudPropVsBackPropHist}, we show the relationship between cloud total pressure and background total pressure, i.e. summing magnetic plus thermal pressure. On average, clouds at all distances are slightly under-pressurised with respect to their surroundings, with a greater contrast with respect to the background layer in comparison to the intermediate layer. A trend with distance is apparent, whereby distributions corresponding to smaller distances are skewed towards more negative values. Similar trends are observed in the contrast distributions with respect to the background layer, albeit at slightly more enhanced values. In addition to a horizontal offset of the peaks, the distributions corresponding to the background layers are broader than their intermediate layer counterparts, at all distances. 

Although clouds, on average, are only weakly under-pressurised with respect to their surroundings when total pressure is considered, the pressure contrast is more striking when only the thermal component is considered. To illustrate this we show the distributions of thermal pressure contrasts between clouds and their surroundings in the centre-right panel of Figure~\ref{fig:cloudPropVsBackPropHist}. The median contrasts are much larger, peaking at roughly $-0.45$ dex and $-0.6$ dex with respect to the intermediate and background layers, respectively, in the inner halo, and with a similar distance trend as the previous panel. It is clear that magnetic pressure is an important component for these clouds, modifying cloud-wind interaction, and without which they would need to contract further to reach pressure equilibrium \citep{sparre2020,nelson2020,li2020}.

In the lower-left panel of Figure~\ref{fig:cloudPropVsBackPropHist}, we show PDFs of cloud-background metallicity contrast. On average, clouds are slightly more enriched than both their intermediate and background layers, at all distances. This may naturally arise since gas with metals cools faster (i.e. through metal line cooling), and is hence more likely to be cold. For both the intermediate and background layers, a very weak trend of this metallicity contrast with distance is present: in the inner halo, the median difference of cloud metallicity with respect to both the surrounding layers is overall negligible ($\sim 0.01$ dex) whereas it is somewhat larger in the central and outer halo (by $\sim 0.04$ and 0.07 dex). Additionally, at all distances, the distributions of contrasts with respect to the background are more broad in comparison to contrasts with respect to the intermediate layer, i.e. clouds are more homogeneous with their immediate surroundings than with gas further away, as one would intuitively expect. All distributions are further skewed towards positive contrasts, i.e. clouds are much more likely to be over-metallic with respect to their surroundings than otherwise.

Finally, the lower-right panel of Figure~\ref{fig:cloudPropVsBackPropHist} shows the kinematic connection between cold clouds and the ambient media. In particular, we show the distribution of cloud-background relative radial velocity. On average, clouds at all distances are (weakly) inflowing with respect to their surrounding gas layers. Distance trends are again noticeable, with clouds at smaller galactocentric distances inflowing faster than their surroundings as compared to more distant clouds. Radial velocity contrasts with respect to the intermediate layer peak at few km s$^{-1}$ in all three regions of the halo, but the widths of these distributions are large: tens of km/s, and are greater at smaller galactocentric distances. With respect to the background layer, a distance trend is noticeable in both the median and width of distributions: a median contrast of roughly $-25$ km s$^{-1}$ in the inner and central regions of the halo reduces to roughly $-5$ km s$^{-1}$ in the outer halo, in addition to widths of distributions being larger at smaller galactocentric distances. 

Such velocity differences are a combined result of inflow/outflow motion of the cold cloud and the inflow/outflow of the background. There are four different possibilities:

\begin{itemize}
    \item Cloud outflowing and intermediate layer (background layer) outflowing: This accounts for $\sim 26$ per cent of all cases. The median contrast of radial velocity in this case is roughly $-7.2$ ($-15.4$) km s$^{-1}$, i.e. when both clouds and the surroundings are outflowing, clouds are outflowing slower with respect to the surrounding. This suggests ongoing acceleration. 
    
    \item Cloud inflowing and intermediate layer (background layer) outflowing: This accounts for $\sim 4$ per cent ($11 $ per cent) of all cases. The median contrast of radial velocity here is approximately $-24.7$ ($-60.6$) km s$^{-1}$, i.e when clouds are inflowing amidst outflowing surrounding gas, there is a large velocity contrast present. These relatively rare cases are likely accreting clouds which are being hit by outflowing galactic-scale winds.
    
    \item Cloud inflowing and intermediate layer (background layer) inflowing: This accounts for $\sim 69$ per cent ($62$ per cent) of all cases. The median value of radial velocity contrast in this case is roughly $-10.9$ ($-22.9$) km s$^{-1}$, i.e. when both clouds and the surrounding are inflowing, clouds are inflowing faster than their surroundings. This suggests ballistic acceleration enabled by cloud overdensities $\chi > 1$. 
    
    \item Cloud outflowing and intermediate layer (background layer) inflowing: This accounts for $\sim 0.5$ per cent ($1.1 $ per cent) of all cases. The median radial velocity contrast here is $\sim 10.8$ ($21.5$) km/s, and this is the only case in which clouds, on average, are outflowing faster than their surroundings. These rare cases reflect outflow-driven, or in-situ outflow formed clouds, which are no longer comoving with any bulk outflow and so can interact with the ambient CGM.
\end{itemize}

The distribution of radial velocity contrasts can thus be summarised as follows: (a) Most clouds ($\gtrsim90$ percent) are flowing in the same direction as their surroundings and with typically small velocity contrasts (few km/s). This reflects the peak of the radial velocity contrast distributions.
(b) Clouds flowing in the opposite direction with respect to their surroundings typically have much larger velocity contrasts (few tens of km/s), and populate the tails of the radial velocity contrast distributions. Although this needs to be investigated in more detail, we speculate that the latter is less common because cold clouds are possibly destroyed as a result of enhanced instabilities when there is a large velocity contrast with respect to the surrounding, for example, as a result of increased mass loss driven by the KH instability \citep[e.g.][]{sander2021}. However, if the cold clouds that we identify are predominantly related to cold gas stripping of satellites, the former case would naturally be more common, since cold gas that is stripped falls with roughly the same velocity as the satellite shortly after being stripped.

\begin{figure*}
\centering 
\includegraphics[width=8cm]{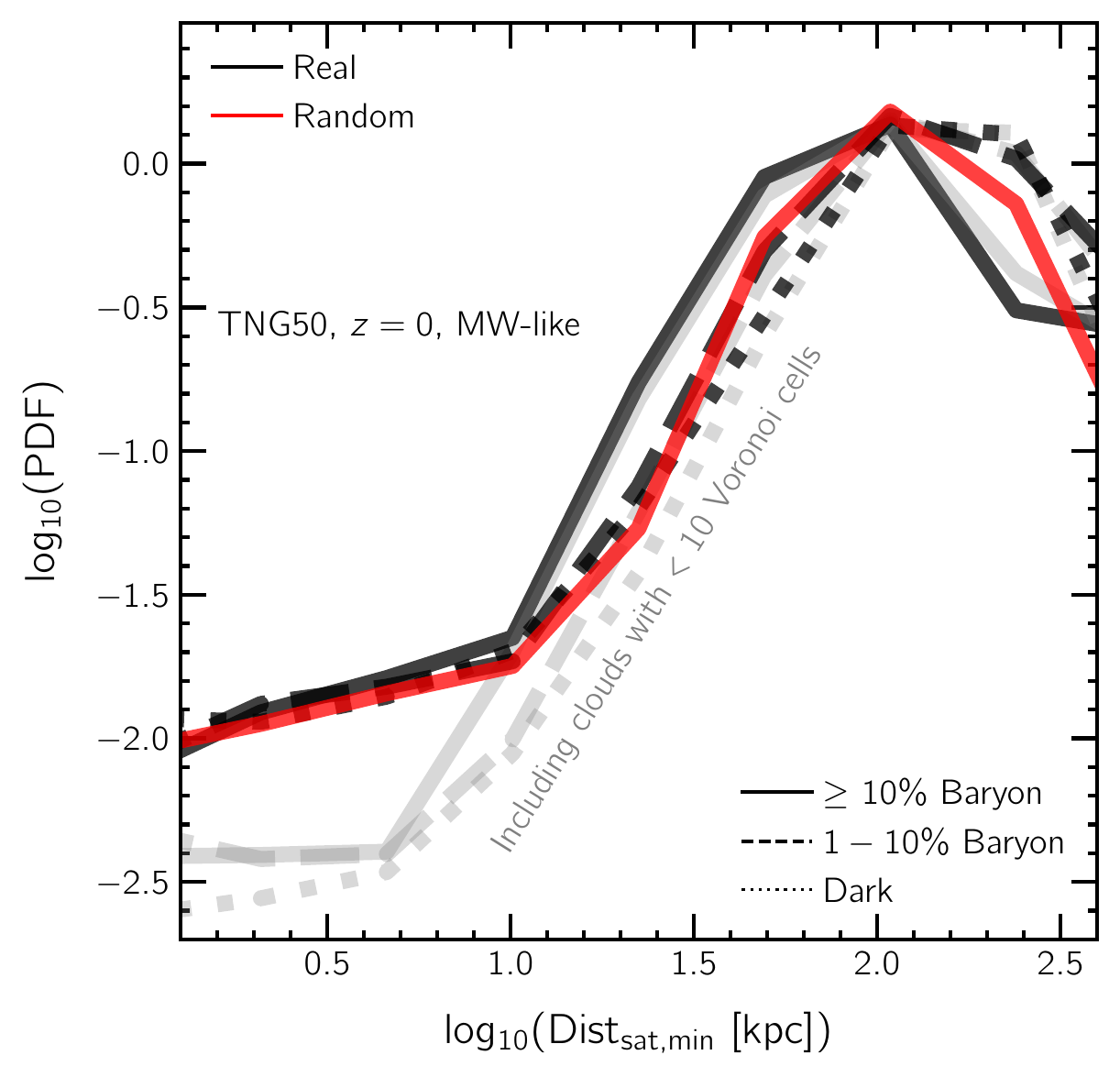}
\includegraphics[width=8cm]{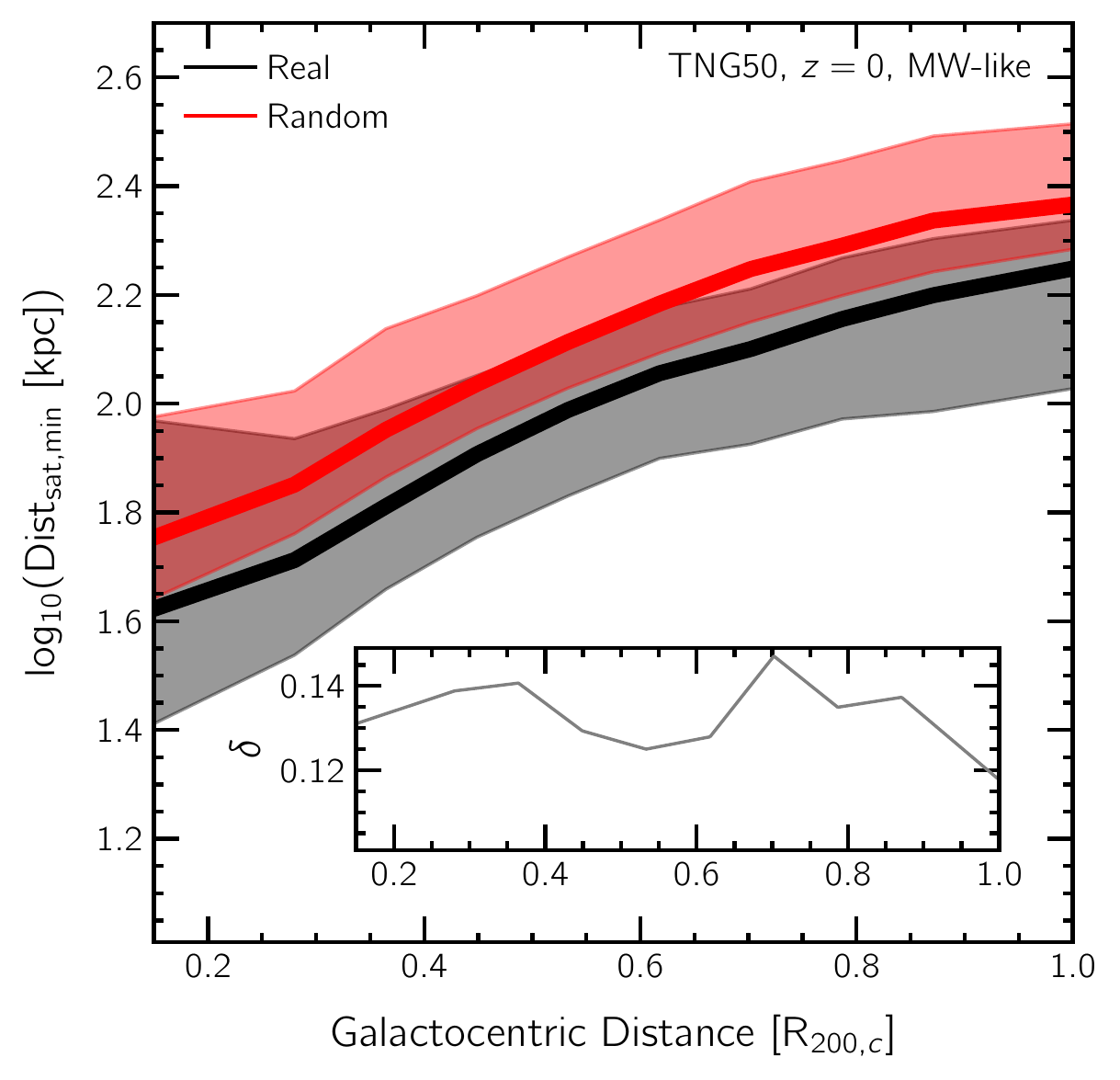}
\caption{The clustering of cold clouds around satellite galaxies in the CGM of TNG50 MW-like galaxies. In the left panel we show the distribution of distances between clouds and their nearest satellite (black lines). We split each PDF into different categories based on the (current) baryon fractions of those satellites: $> 10$ per cent, $1 - 10$ per cent and $0$ per cent in solid, dashed and dotted curves, respectively. As a comparison point, we show the expected distribution after randomizing the position of the nearest satellite (red curve, see text). Clouds are more frequently found close to satellites than would be expected in the random case, this being the case at least for satellites with higher baryon fractions. The right panel shows the relation between the distance to the nearest satellite with a baryon fraction $> 10$ per cent (y-axis) and the galactocentric distance of the cloud (x-axis), for both the real and the random case. The solid curves show the median, while the shaded bands correspond to the 16$^{\rm{th}}$ and 84$^{\rm{th}}$ percentile regions, of the stacked sample of clouds across all 132 MW-like galaxies. The inset shows the difference between the two median curves, i.e. random minus real. At all distances, on average, the correlation between positions of clouds and their nearest satellites is stronger than what is expected if their relative positions were random.}
\label{fig:cloudSpatClustr_1}
\end{figure*}

\subsection{Spatial clustering of CGM cold clouds}\label{sec:clouds_spatial_clustering}

As we have seen (Figure~\ref{fig:cloudPosSatPos}), cold gas clouds in the CGM of TNG50 MW-like galaxies are not distributed uniformly throughout the halo, but rather appear to be clustered in certain locations, often near satellite galaxies or tails of their stripped gas.

We quantify this phenomenon in Figure~\ref{fig:cloudSpatClustr_1}, where we study the minimum distance between a cold cloud and its nearest satellite.\footnote{Note that clouds related to gas that has been stripped in the distant past would naturally end up with relatively large distance values, despite physically originating from the corresponding satellite. In the future, we will overcome this limitation by using the Monte Carlo tracers to first identify the associated satellite, and then estimate the closest distance to the past trajectory of that satellite.} The left panel shows PDFs of the distance of clouds to their nearest satellite galaxy. As before, we construct individual PDFs for each halo, and then the median across haloes.

We consider three types of satellites, based on their (current) baryon fractions, i.e. ratio of baryon mass to total subhalo mass: $>10$, $1-10$ and $0$ per cent in black solid, dashed and dotted curves, respectively. Since the first category of satellites are typically less frequent ($3-4$ on average per halo) in comparison to the latter two categories, we down sample the number of the latter two to avoid any biases. We do so through the following procedure: if a halo contains $N$ satellites with baryon fraction $> 10$ per cent, we randomly select $N$ satellites of the latter two categories from the available pool. To make sure that this random-picking does not bias our results, we repeat the process 100 times for each galaxy.

Overall, we find that cold clouds tend to be closer to baryon rich satellites than either dark satellites or (relatively) baryon poor satellites, suggesting a physical i.e. origin link. A vertical offset in the PDFs is seen at $\lesssim 100$\,kpc as one transitions towards lower values of baryon fractions of satellites. Clouds are thus more likely to lie close to satellites with larger baryon fractions. 

A similar observation of cloud-satellite clustering exists in the Milky Way halo, where there is a high concentration of HVCs around the Magellanic Stream, and so also close to the Large and Small Magellanic Clouds \citep{moss2013}. Idealized simulations of gas-rich dwarf galaxy stripping also show similar signatures \citep{mayer2006}.

To confirm that this correlation between positions of cold clouds and satellites with large baryon fractions is robust, we carry out a random shuffling experiment, as follows. Once the nearest satellite to a particular cloud is identified, we randomize the position of that satellite, and re-compute the minimum distance between the cloud and the randomized position of the satellite. To avoid any biases, we repeat this procedure 100 times. If the position of a cloud with the associated satellite was truly random, this procedure would have minimal effect. However, if the connection between a cloud and a satellite is `real', the randomization would erase any signature. We show the median PDF corresponding to the random case in red. A clear offset is once again visible, suggesting that a physical correlation is indeed present. 

In the right panel of Figure~\ref{fig:cloudSpatClustr_1}, we show how the distance to the nearest baryon-rich satellite satellite depends on the galactocentric distance of the cloud. The solid lines show the median, while the shaded regions correspond to the 16$^{\rm{th}}$ and 84$^{\rm{th}}$ percentile regions. In black, we show the trend corresponding to the case emerging from the simulation (`real' case). A median distance to the nearest baryon-rich satellite of $\sim40$\,kpc at the inner boundary of the CGM (i.e. $0.15 \RVIR$) increases to $\sim80$\,kpc at $0.5 \RVIR$, and further to $\sim 180$\,kpc at the virial radius. The width of the percentile regions simultaneously decreases, from $\sim 0.5$\,dex at $0.15 \RVIR$ to $\sim 0.3$\,dex at the virial radius. 

Intuitively, we expect a similar qualitative trend even for clouds/satellites distributed randomly, since shells (of equal width) at smaller galactocentric distances have smaller volumes. To normalize out this volume effect, we also include the corresponding relations for the random case. As expected, a similar trend with distance is seen, although the red median is clearly offset vertically above the black median curve. The difference between these two medians is shown in the inset, which estimates the true strength of the radial dependence of clustering, i.e. with the volume-scaling effects removed. The offset is rather independent of distance, varying between $\sim 0.12$ and $0.15$\,dex. Thus, at all distances, a weak over-correlation is seen between the positions of clouds and $>10$ per cent baryon fraction satellites with respect to the random case.

While the main results of this panel are largely unaffected by the lower-limit threshold for the minimum number of cells per cloud, we mention a subtle difference that is present for the case where clouds with less than ten member cells are included: the corresponding distributions are shown in gray curves. While the gray and black distributions merge for Dist$_{\rm{sat,min}}$ values larger than $\sim 10$\,kpc, a vertical offset is present in the solid and dashed curves with respect to the dotted one at smaller distances. Although the difference is small, we suspect that it could correspond to gas that has been freshly stripped, and is hence present in the distributions corresponding to baryonic satellites, but absent in the dark case. While this could simply be a result of poor resolution, it may also be the case that these tiny (unresolved) clouds act as seeds of dense, cold gas that trigger thermal instability, eventually giving rise to larger clouds as gas condenses around them \citep{nelson2020,dutta2022}.

\begin{figure*}
\centering 
\includegraphics[width=8.05cm]{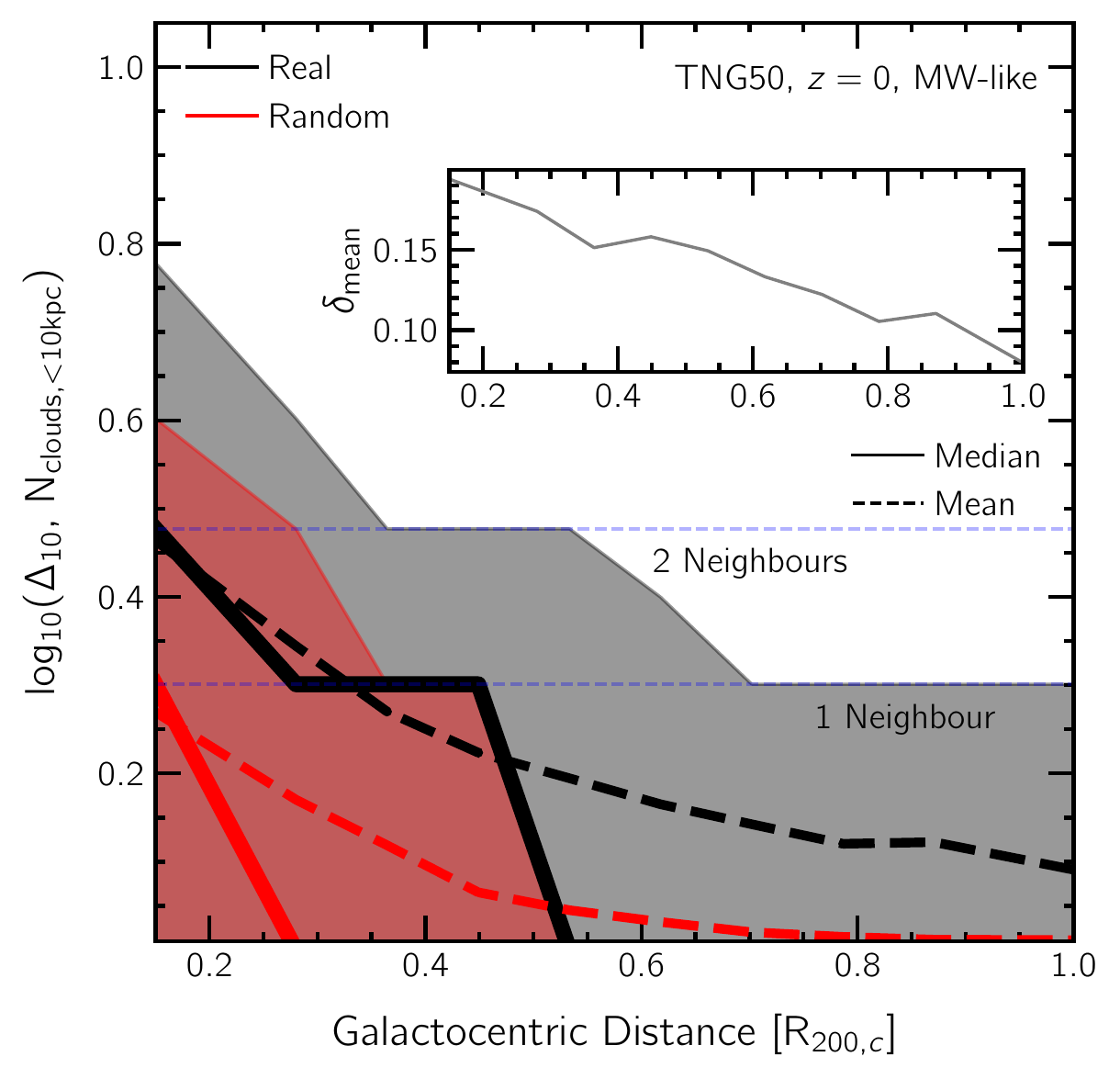}
\includegraphics[width=8cm]{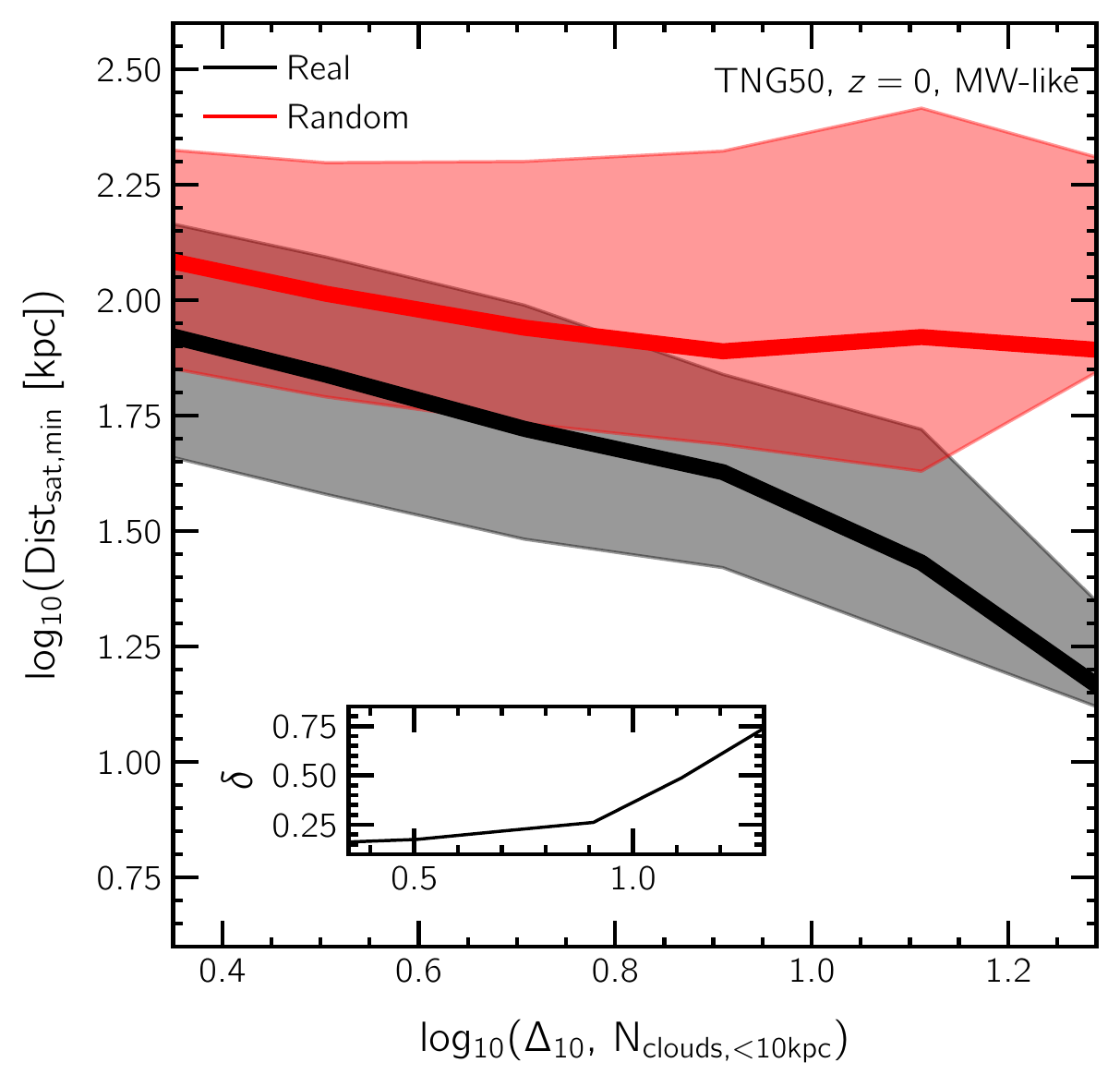}
\caption{The spatial clustering of cold clouds around other cold clouds in the CGM of TNG50 MW-like galaxies. The left panel shows the trend between $\Delta_{10}$, the number of clouds within spheres of radius 10 kpc centered on each identified cloud, and galactocentric distance. Solid curves correspond to median values, dashed curves to mean values, and shaded regions to 16$^{\rm{th}}$ and 84$^{\rm{th}}$ percentiles. Black curves show the actual outcome of the simulation, whereas the red curves show what a random (shuffled) spatial distribution would look like. The inset shows the difference between the two mean values. In the right panel, we show the trend between $\Delta_{10}$ and the minimum distance of clouds to the nearest satellite with baryon fraction $>10$ per cent, for both the real case (black), and the case where satellite positions are randomized (red). The inset shows the difference between the two medians. Clouds are typically over-clustered around other clouds with respect to what a random scenario would predict, with stronger over-clustering at smaller galactocentric distances. Higher values of $\Delta_{10}$ are seen when clouds lie closer to satellite galaxies with ``high'' baryon fractions.}
\label{fig:cloudSpatClustr}
\end{figure*}

We next turn to the possibility that cold clouds may be clustered around other cold clouds, rather than being distributed uniformly throughout the halo. It is believed that such clustering can increase the longevity of clouds through the process of drafting \citep{williams2022}. We quantify this clustering through the $\Delta_{10}$ parameter, which we define as follows: for every identified cloud, $\Delta_{10}$ is the number of clouds that lie within a sphere of radius 10kpc, including itself. A value of one thus implies that there are no neighbouring clouds within this sphere, a value of two corresponds to one neighbour, and so on. As extended clouds are much less frequent than smaller ones, in this statistical approach we neglect the issue that arises because clouds are in fact extended objects.

The left panel of Figure~\ref{fig:cloudSpatClustr} shows the trend of $\Delta_{10}$ with galactocentric distance. Solid curves correspond to median values, dashed curves to mean values, and shaded regions to 16$^{\rm{th}}$ and 84$^{\rm{th}}$ percentiles. We begin by discussing the black curves, which represent the actual outcome from the simulation. A median value of 2 neighbours (i.e. log$_{10}(\Delta_{10}$) = log$_{10}(3)$) in the innermost regions of the halo ($\lesssim 0.3 \RVIR$) reduces to 1 neighbour (i.e. log$_{10}(\Delta_{10}$) = log$_{10}(2)$) between $0.25 \RVIR$ and $0.5 \RVIR$, and to zero neighbours at farther distances. The mean, however, does not portray such a `step-like' behaviour, and shows a steady monotonic decrease of log$_{10}(\Delta_{10})$ with distance, reducing from $\sim 0.5$ at $0.15 \RVIR$ to $\sim 0.1$ at the virial radius, i.e. at all distances, the mean number of neighbours is more than zero, and is greater at smaller galactocentric distances.

As before, such a qualitative trend with distance is expected even for a random distribution of clouds, due to available volume decreasing towards the halo center. To remove this effect, we randomize the positions of clouds while keeping their radial number density profile fixed, and re-calculate $\Delta_{10}$ for the randomized positions. The corresponding trend is shown in red. If the positions of clouds with respect to other clouds was already random, this procedure would not have a significant impact. However, in case clouds are truly clustered around their neighbours, a difference would emerge. Indeed, in both the mean and the median, the randomized scenario shows a smaller value of $\Delta_{10}$ than the true case, at all distances. The inset shows the difference between the two mean values ($\delta_{\rm{mean}}$), characterizing the strength of the true radial dependence of clustering. Over-clustering with respect to the random case is strongest at smaller galactocentric distances: a $\delta_{\rm{mean}}$ of $\sim 0.19$ at $0.15 \RVIR$ reduces to $\sim 0.15$ at $0.5 \RVIR$, before dropping steeply to $\sim 0.08$ at the virial radius.

While we use the $\Delta_{10}$ metric to study clustering, other statistics are equally well suited. For example, the two-point correlation function, or e.g. the distance to the tenth nearest cloud, as a measure of over-clustering. We have considered both and they provide qualitatively similar results, demonstrating an over-clustering of cold clouds with respect to the random scenario.

In the right panel of Figure~\ref{fig:cloudSpatClustr}, we show that $\Delta_{10}$ is linked to the minimum distance to the nearest satellite with a baryon fraction of $>10$ per cent. Median values are shown with solid curves, and 16$^{\rm{th}}$ and 84$^{\rm{th}}$ percentiles with shaded regions. The black curve shows the true signal, while the red shows a test where the position of satellites are randomized, as discussed above. The black median curve shows a sharp monotonic drop with increasing $\Delta_{10}$: when a cloud has only one neighbour within 10 kpc, the average distance to the nearest baryon-rich satellite is $\sim 100$kpc. When a cloud instead has $\gtrsim 10$ neighbours in close proximity, the median distance to the nearest baryon-rich satellite drops to $\lesssim 20$kpc. 

The random case portrays a qualitatively similar trend, albeit with an offset of $\sim 0.2$ dex at low values of $\Delta_{10}$, and a shallower drop towards higher values of $\Delta_{10}$. Most importantly, the $\Delta_{10}$-trend of the offset between the two median curves is shown in the inset. An offset of $0.2$dex at log$_{10}(\Delta{10}) \sim 0.3$ rises sharply to $0.75$dex at log$_{10}(\Delta{10}) \sim 1.4$. Thus, when clouds are strongly clustered with neighbouring clouds, they are more likely to lie close to a $> 10$ per cent baryon fraction satellite, as opposed to being randomly positioned with respect to such satellites.

\begin{figure*}
\centering 
\includegraphics[width=8.4cm]{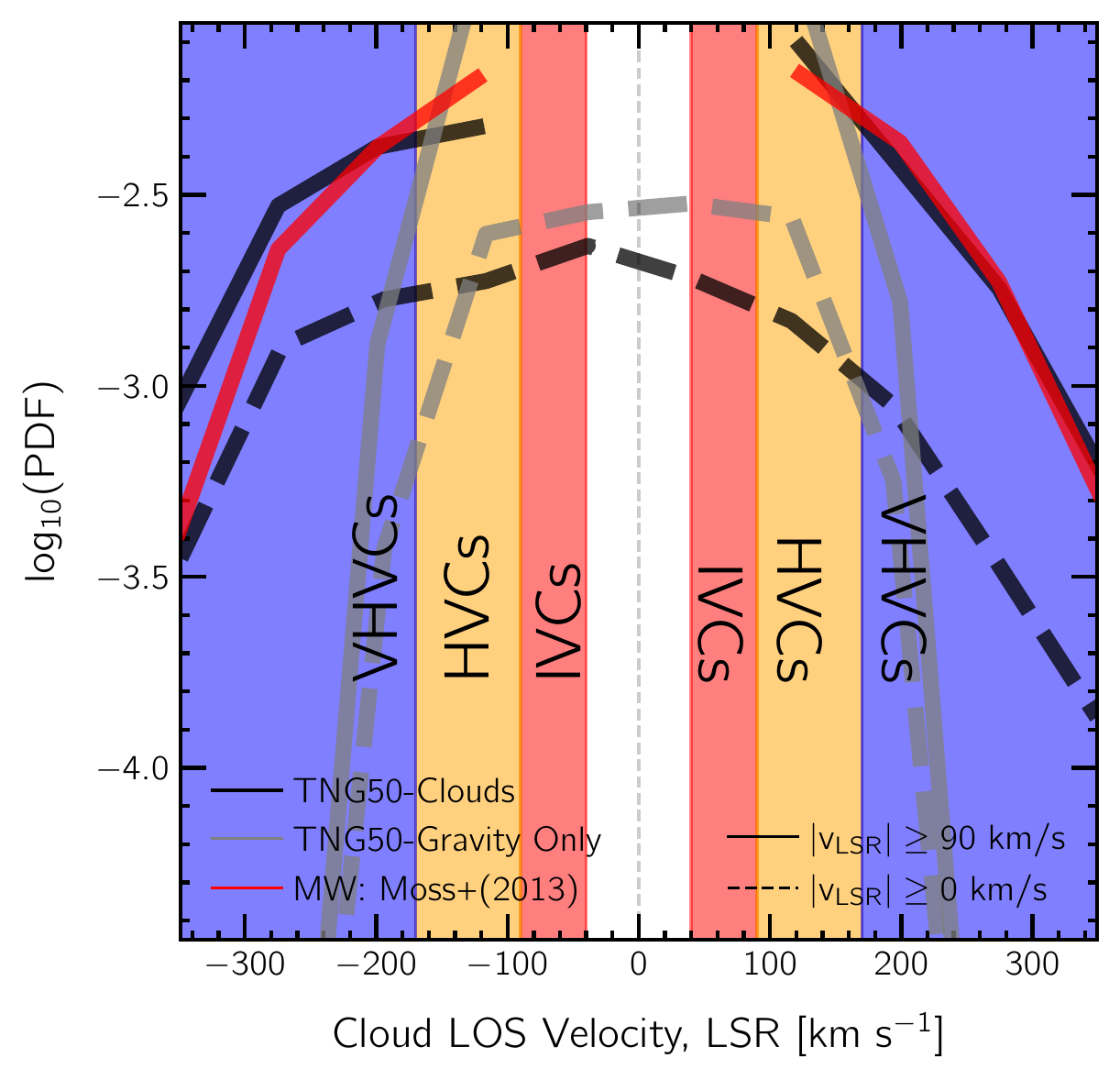}
\includegraphics[width=8cm]{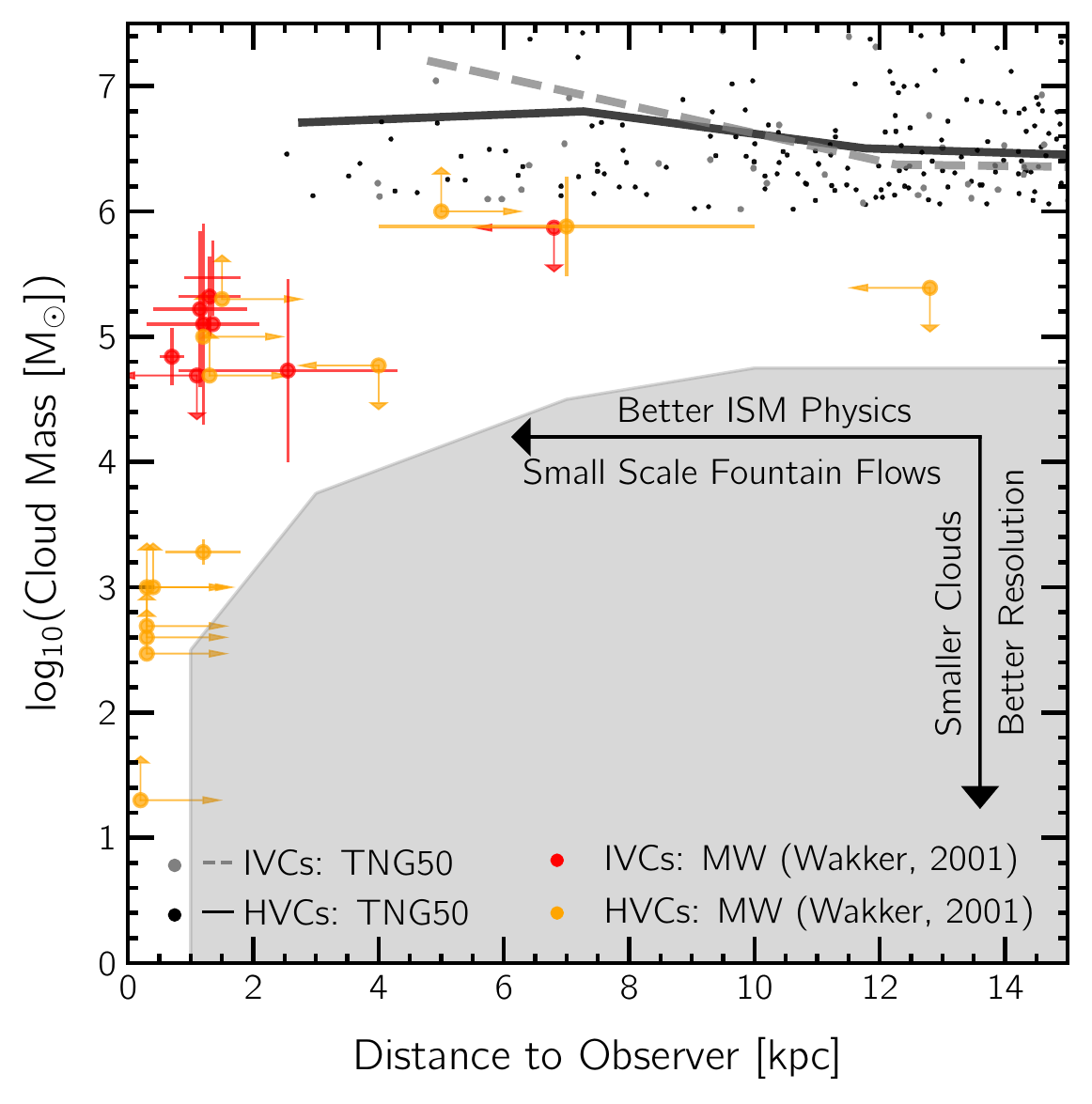}
\includegraphics[width=16cm]{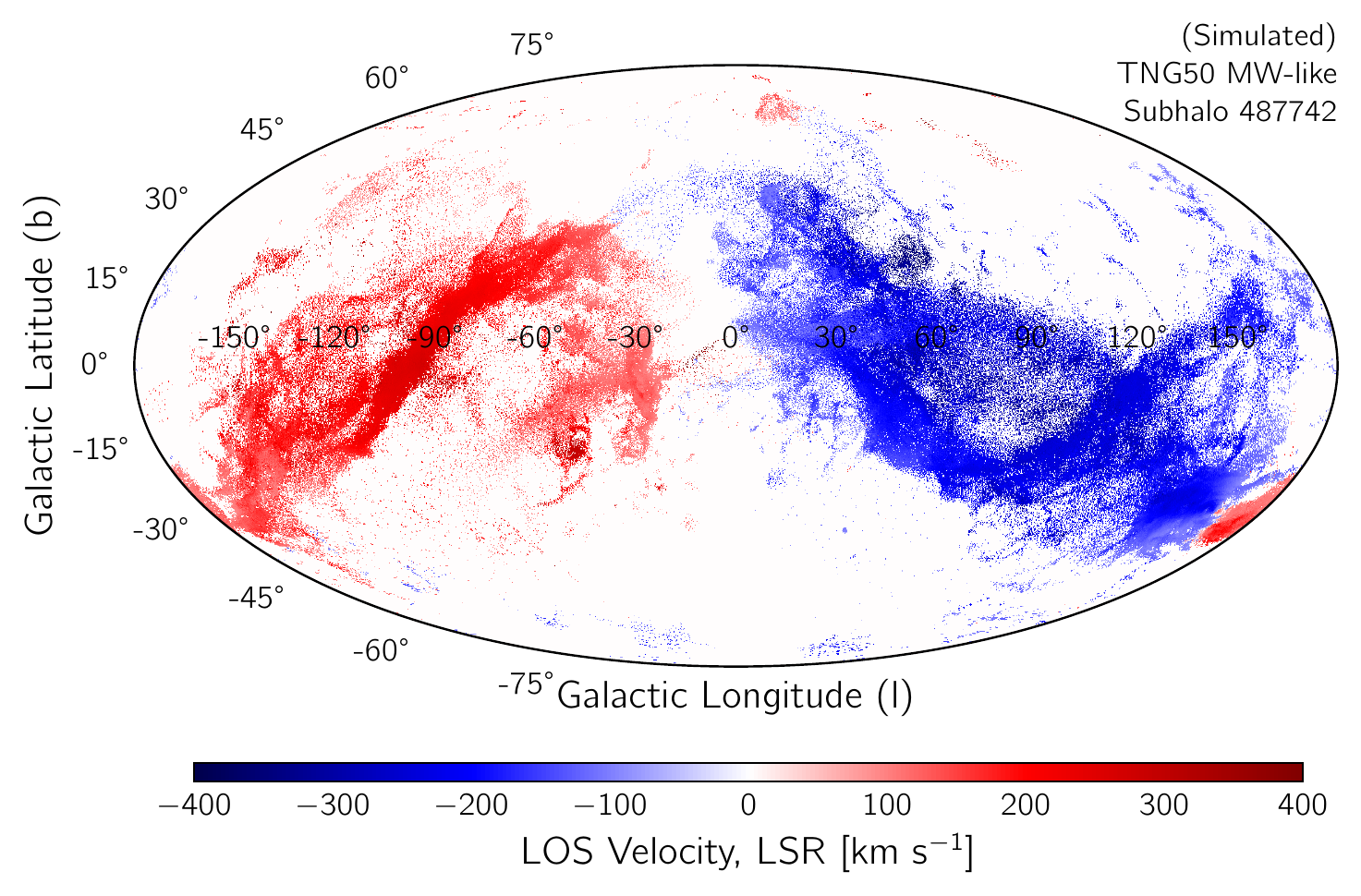}
\caption{Comparison between the clouds in TNG50 MW-like galaxies with observational data of the real Milky Way. In the top-left panel, we show median PDFs for the line-of-sight (LOS) velocity in the frame of reference of the local standard of rest (LSR). Shaded bands in the background correspond to commonly used definitions for clouds. Red curves correspond to distributions from a catalog of observed HVCs in the Milky Way halo (\protect\citealt{moss2013}), while black curves show the equivalent distributions from TNG50 across all 132 MW-like galaxies. We also include the distribution for hypothetical test particles (gray) under purely gravitational motion, demonstrating that the kinematics of HVCs are more complex than gravity alone. In the top-right panel, we show the relation between cloud mass and observer-centric distance. Orange and red points show observational data (\protect\citealt{wakker2001}), while black and gray points and median curves are from TNG50, for HVCs and IVCs, respectively. The bottom panel shows an all-sky Aitoff projection of a TNG50 MW-like galaxy (subhalo 487742), from a hypothetical observer at the solar location, with colors corresponding to the line-of-sight velocity of cold gas, in the frame of the LSR. The distribution of neutral hydrogen in the CGM has a complex morphological and kinematic structure.}
\label{fig:obs_plots}
\end{figure*}

\begin{figure*}
\centering 
\includegraphics[width=18cm]{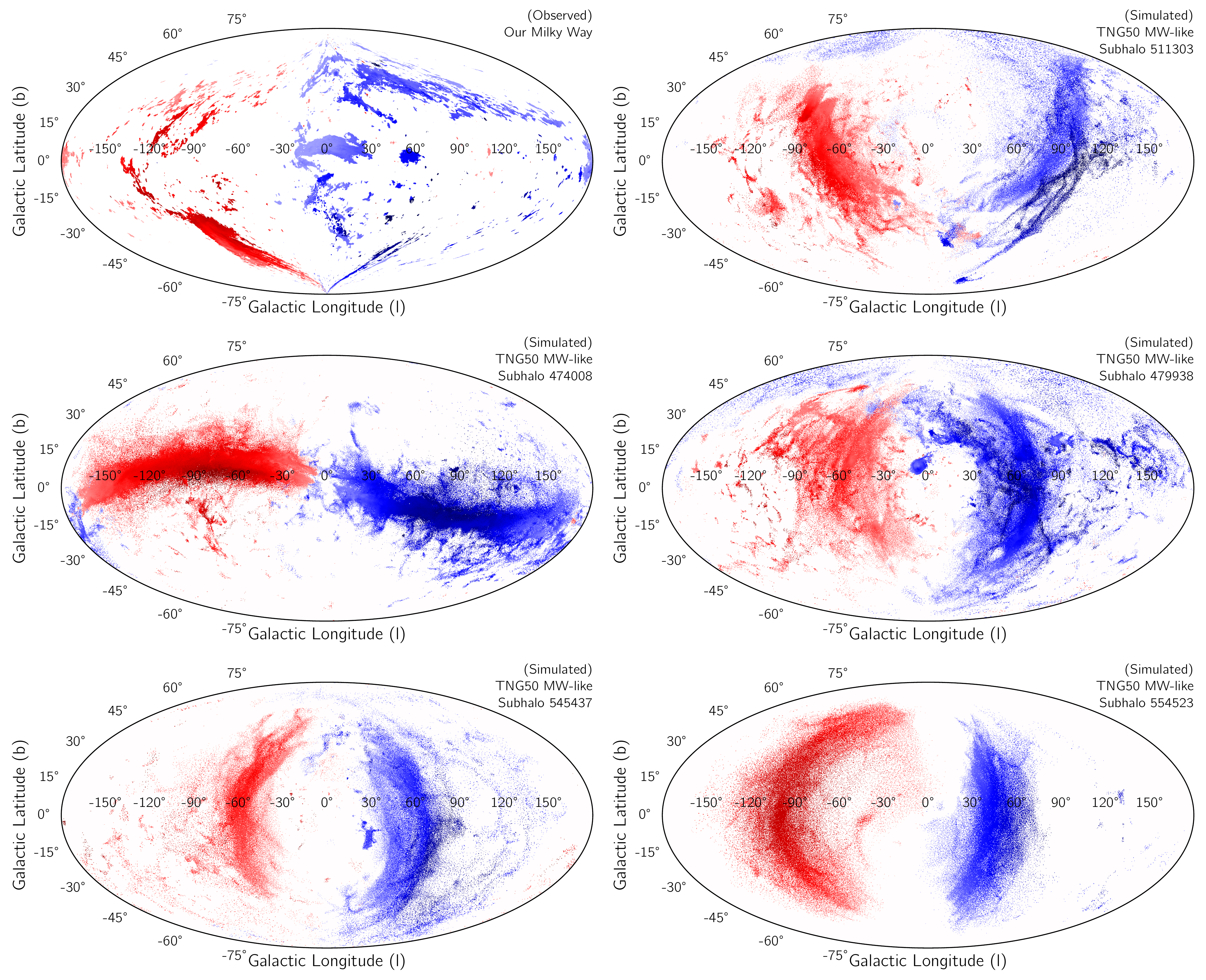}
\includegraphics[width=8cm]{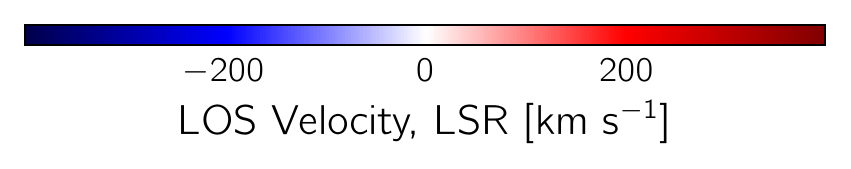}
\caption{Aitoff projections of cold gas, colored by line of sight velocity, similar to the bottom panel of Figure~\ref{fig:obs_plots}. In the top-left panel, we show the all-sky projection from the HI4PI survey (\protect\citealt{westmeier2018}) for the true Milky Way halo, while all other panels correspond to projections from five randomly selected MW-like galaxies from our TNG50 sample. The most interesting is the top-right panel (subhalo 511303), which contains a SMC/LMC-like pair. Similar to the top-left panel, a Magellanic-like stream around $[l,b] \sim [0, -60]$ deg is present in this case. Overall, the distribution of cold gas through the halo is both spatially and kinematically complex, as well as diverse.}
\label{fig:aitoff}
\end{figure*}

\subsection{Comparison with observations of the Milky Way}\label{sec:comp_obs}

We conclude our investigation with a number of direct comparisons with observed data of IVCs and HVCs in the Milky Way halo. This important connection is enabled by the cosmological context of the TNG50 MW-like galaxies, and is unavailable in single cloud and other idealized numerical simulations.

In what follows, we place a hypothetical observer at a random point in the galactic plane, at a distance of $8.34$~kpc away from the galactic centre. This observer is considered to be in perfect circular motion around the galactic centre, at a velocity of $240$~km s$^{-1}$. This observer is consistent with the known solar location and motion in the real Milky Way \citep{reid2014}. Since observations do not enforce a minimum radial distance when identifying clouds, we here relax our lower limit for the inner boundary of the CGM, i.e. we include all clouds present within the virial radius of the halo, barring the one massive cloud that is the galaxy itself. 

In the top left panel of Figure~\ref{fig:obs_plots}, we show PDFs of the line of sight velocity of cold clouds. The different colored regions signify common definitions used to classify clouds: IVCs are those with (absolute) line of sight velocities in the range $40 - 90$ km s$^{-1}$, HVCs with $90 - 170$ km s$^{-1}$, and VHVCs with $> 170$ km s$^{-1}$ \citep[e.g.][]{lehner2022}, although some authors refer to all clouds with (absolute) line of sight velocities $> 90$ km s$^{-1}$ as HVCs \citep[e.g.][]{wakker2001}. The red curves show the PDF of a sample of HVCs of the Milky Way presented in \cite{moss2013}. The solid black curve shows the median PDF of HVCs across the whole TNG50 MW-like sample of simulated galaxies. Overall, the agreement is striking. Both curves indicate that the abundance of clouds is smaller for higher velocity clouds. Note that the TNG50 result is the stacked outcome, averaging across all galaxies of our sample, and these have a diversity of properties including stellar disk lengths (see \textcolor{blue}{Pillepich et al. in prep}). In the black dashed line, we show the median PDF of all clouds in TNG50, irrespective of velocity. We predict that the abundance of `low velocity' clouds with (absolute) line of sight velocities $< 40$ km s$^{-1}$ -- not generally accessible in observations -- is roughly independent of velocity, and that these clouds are more abundant than higher velocity clouds.

To check if the motion of these clouds is dominated by gravity, or not, we compare the velocity distribution of clouds to that of hypothetical test particles whose motion is purely determined by gravity. To do so, for each cloud, we compute the (circular) velocity that is required at that distance for the centrifugal force to perfectly balance the force of gravity. The PDFs of these test particles are shown in gray: those consistent with HVC velocities as solid curves, and for all velocities in dashed curves. While the shape of the gray dashed curve is similar to that of the black dashed curve in the velocity range $\sim [-90, 90]$ km s$^{-1}$, i.e. clouds that are not HVCs (or VHVCs), the two curves diverge at higher velocities. That is, gravity alone cannot account for the high velocity tails, suggesting that further astrophysical processes are relevant for the kinematics of HVCs.

In the top right panel of Figure~\ref{fig:obs_plots}, we show the relation between cloud mass and the (three-dimensional) distance to the observer. The different scatter points show estimates of IVCs (red) and HVCs (orange) from \cite{wakker2001} for the Milky Way. Note that most of these points are either lower or upper limits, which we denote with arrows in the relevant direction. Since all these clouds were observed through their HI emission, \cite{wakker2001} use factors of $1.2$ and $1.39$ to account for the masses of ionized hydrogen and helium respectively, to arrive at a better estimate of total mass, from their initially inferred HI mass. However, more recent studies suggest that the ionized component of clouds may account for a larger mass fraction \citep[e.g.][]{lehner2011}, and thus the masses quoted by \cite{wakker2001} are likely underestimated. As is, most clouds in the sample have a mass $\lesssim 10^5 \rm{M_\odot}$, i.e. below the resolution limit of TNG50. 

Although these data points follow the expected $M \propto d^2$ dependence \citep[e.g.][]{wakker1997}, we suspect this to be largely due to the prevalence of lower limits for the distance estimates of a large fraction of clouds, especially for less massive clouds ($\lesssim 10^5 \rm{M_\odot}$). This artificially results in too small cloud mass estimates. If one were to instead assume a uniform distribution of distances in the range $\sim [5,12]$kpc, as motivated by absorption line measurements which are insensitive to mass of clouds \citep[e.g.][]{lehner2012, lehner2022}, a different mass distribution would arise. Indeed, for those clouds more massive than $10^5 \rm{M_\odot}$ in \cite{wakker2001}'s sample, where distance estimates are a mix of lower-limits, upper-limits and tighter constraints, there seems to be no strong dependence of mass with distance, at least within the distance brackets available.

For comparison, we show TNG50 IVCs (HVCs) with gray (black) points, and their median with the gray (black) curve. Consistent with the few observational constraints available, TNG50 predicts a weak dependence on distance for clouds above $\sim 10^6 \rm{M_\odot}$, and a very similar relation for both IVCs and HVCs, out to distances within which clouds in the Milky Way are typically observed. A noteworthy point from this plot is the dearth of TNG50 clouds at small distances. While this is likely because all cold gas in this region is contiguous with the galactic disk, it is possible that the lack of these `small-scale fountain flows' is a limitation of the simplified stellar feedback driven galactic-wind model of TNG (Section~\ref{methods}; \citealt{springel2003}). Further exploration with alternate, more explicit stellar feedback models \citep[e.g.][]{smith2018,hopkins2020,hu2022} would be be essential to comment on small distance clouds. We demarcate the large region inaccessible to TNG50 with the gray region in the lower right corner. Moving to small cloud masses simply requires higher numerical resolution, while moving closer to the disk-halo interface requires more sophisticated models for ISM and stellar feedback physics.

In the bottom panel of Figure~\ref{fig:obs_plots}, we show an all-sky map of gas line-of-sight velocity, including only high-velocity ($> 90$ km s$^{-1}$) cold gas (T $<10^{4.5}$K) in emission. We show a single TNG50 MW-like galaxy, the same halo from Figure~\ref{fig:cloudPosSatPos}, as seen by our hypothetical observer. To be as realistic as possible, we include all halo gas within the virial radius, including gas that is gravitationally bound to satellites, since sky maps of the Milky Way halo contain such components. Colors indicate the velocity at which gas is moving along the line of sight to the observer.

A gallery of such projections are shown in Figure~\ref{fig:aitoff}, where the top-left panel is observational data of the real Milky Way, from the HI4PI survey \citep{westmeier2018}. The other five panels are five randomly selected galaxies from our TNG50 WW-like sample. The most interesting is the top-right panel, which shows a halo that contains a SMC/LMC-like pair. This map exhibits a degree of qualitative similarity to the true data: along with a noticeably patchy distribution of gas throughout the halo, a Magellanic-like stream at $[l,b] \sim [0, -60]$ deg is present. Although all the other TNG50 all-sky projection lack a Magellanic-like stream, a trend in colors is apparent across all these maps, with gas at negative longitudes preferentially outflowing, while gas at positive longitudes is preferentially inflowing. Gas distributions are clearly unique in each TNG50 halo, highlighting the diversity across the sample.

\section{Summary and Conclusions}\label{summary}

In this paper, we have studied the existence, distribution, and physical properties of cold, dense clouds of gas in the circumgalactic medium (CGM) of a sample of 132 Milky Way-like galaxies in the TNG50 simulation at $z=0$. Our motivation to study such objects stems from the plethora of open questions surrounding high velocity clouds (HVCs) in the real Milky Way halo. TNG50 offers a combination of resolution and volume to begin exploring such clouds in a cosmological context, over a wide sample of galaxies, bridging the gap to small-scale, idealized numerical simulations of cold cloud evolution and survival. We summarise our main findings as follows:

\begin{enumerate}
    \item MW-like galaxies in TNG50 typically contain of order one hundred (thousand) reasonably (marginally) resolved cold clouds in their gaseous haloes. While the number of clouds shows no significant trend with the stellar mass of the galaxy, the scatter correlates with the specific star formation rate (sSFR). This suggests that (a) AGN feedback quenches star formation and destroys clouds, or prevents clouds from forming, and/or (b) the flow of cool gas through the circumgalactic medium (CGM) is physically connected to the fuelling of galactic star formation (Figure~\ref{fig:numClouds}).
    
    \item Clouds show a large variation in their mass, although most clouds in our sample have a mass close to $\sim 10^{6}$ M$_\odot$, corresponding to the chosen cloud definition. More massive clouds are larger, with cloud sizes ranging from $\sim$ a few hundred pc to $\sim$ a few 10s of kpc. Smaller clouds tend to be found in the inner halo. Clouds also span a wide range of shapes, and smaller clouds are more spherical than their more massive counterparts (Figure~\ref{fig:cloudBasicProp1}).
    
    \item With respect to cloud properties, most clouds ($\sim 90$ per cent) have sub-solar metallicities. However, clouds with metallicity as high as $\gtrsim 2\, Z_\odot$ exist. Most clouds have $\beta \sim 1$, indicating a balance of thermal and magnetic pressure. Magnetic pressure is larger in $2/3$ of clouds, although most clouds outside the inner halo ($> 0.4 \times \RVIR$) are thermally dominated \citep[in contrast to those in more massive haloes in TNG50;][]{nelson2020}. Clouds are typically inhomogeneous in their metallicity content, and temperature-, density- and pressure-structures; inner inhomogenities in their temperature structure are the most significant (Figure~\ref{fig:cloudBasicProp2}).
    
    \item While most clouds have relatively small radial velocities (of order $\sim$\,10 km s$^{-1}$), clouds tracing fast inflows and fast outflows are both present, and these are more prevalent at smaller galactocentric distances. Across the entire halo, inflowing clouds dominate ($\sim 73$ per cent across all MW-like galaxies). Overall, clouds tend to be dominated by sub-virial rotation. Metallicity correlates strongly with radial velocity: rapidly outflowing cold clouds are the most metal rich, whereas rapidly inflowing clouds are the least enriched, hinting at different physical origins (Figures~\ref{fig:cloudBasicProp2} and \ref{fig:cloudKinProp}).
        
    \item We compare the physical properties of clouds to their surrounding gas, defining local `intermediate' (i.e. interface) and `background' layers. On average, cold clouds in the CGM of MW-like galaxies are more metal rich, denser, cooler, and preferentially inflowing with respective to their backgrounds. However, metallicity and velocity contrasts are small, of order 0.05 dex and 10 km s$^{-1}$, respectively, on average. We find a typical overdensity of $\chi$\,$\lesssim$\,$10$, which is larger at smaller halocentric distances, but much smaller than often assumed in idealized cloud simulations. While most clouds are only slightly under-pressurised with respect to their surroundings when total (magnetic plus thermal) pressure is considered, they are significantly thermally under-pressurized. This suggests magnetic fields may be an important pressure component in cold clouds in the CGM of MW-like galaxies (Figure~\ref{fig:cloudPropVsBackPropHist}).
    
    \item Cold clouds are not uniformly distributed throughout the halo, but are strongly clustered. At all distances, clouds have more neighboring clouds (within 10 kpc) than would be expected for a random distribution. This over-clustering is greater towards the halo center. We also find a clear clustering of cold clouds around satellite galaxies with large ($\gtrsim 10$ per cent) baryon fractions. This suggests a stripping origin for at least part of the cold cloud population (Figures~\ref{fig:cloudSpatClustr_1} and \ref{fig:cloudSpatClustr}).
    
    \item Finally, we qualitatively compare results from TNG50 with observations of HVCs in the Milky Way halo. The observed line-of-sight (LOS) velocity distribution of clouds is remarkably consistent with the average MW-like galaxy in TNG50: HVC abundance drops with increasing velocity. We show that the kinematics of cold clouds are not consistent with gravitational motion alone, suggesting that astrophysical feedback processes influence the motion of cold gas in the CGM. For clouds above $\sim 10^6$ M$_\odot$, no trend of mass with distance is seen in TNG50, which is consistent with the limited number of HVC observations available at this mass range. TNG50 predicts that (currently poorly constrained) `low velocity' clouds are the most abundant, and that their abundance is roughly independent of LOS velocity (Figure~\ref{fig:obs_plots}). 
\end{enumerate}

This work is our first attempt to bridge studies of clouds using idealized, small-scale, controlled numerical experiments including wind tunnel or `cloud crushing' simulations with those using MW-like galaxies realised through large-volume, cosmological, galaxy formation simulations. Building upon the study of \cite{nelson2020} that found large abundances of cold clouds in high-mass TNG50 haloes, we have focused specifically on galaxies that resemble our own Milky Way.

However, even with TNG50 we have here considered physical phenomena right at the edge of available numerical resolution. We cannot yet demonstrate that the abundance of small, low-mass clouds ($\lesssim 100$'s\,pc, $\lesssim 10^5$\,M$_\odot$) is realistic, nor converged. Future work requires simulating the circumgalactic medium of MW-like galaxies at significantly higher resolution, and with currently missing physics such as radiation transport and cosmic ray pressure components. This will enable the study of even smaller-scale structures, while also better resolving the physical properties and evolutionary origins of the HVC-like cloud structures already present in TNG50.

\section*{Data Availability}

The IllustrisTNG simulations, including TNG50, are publicly available and accessible at \url{www.tng-project.org/data} \citep{nelson2019b}. New data products for the Milky Way/M31-like sample are now on the same website ({\color{blue} Pillepich et al. in prep}). The `cosmological cloud catalog' produced and used in this work is publicly released at \url{www.tng-project.org/ramesh23b}. Other data related to this publication is available upon reasonable request. 

\section*{Acknowledgements}

RR and DN acknowledge funding from the Deutsche Forschungsgemeinschaft (DFG) through an Emmy Noether Research Group (grant number NE 2441/1-1). AP acknowledges funding by the DFG -- Project-ID 138713538 -- SFB 881 (``The Milky Way System'', subprojects A01 and A06). RR is a Fellow of the International Max Planck Research School for Astronomy and Cosmic Physics at the University of Heidelberg (IMPRS-HD). RR and DN thank J. Christopher Howk for helpful discussions on the observational aspects of clouds. The authors also thank Ralf Klessen for valuable inputs. The authors thank the anonymous referee for constructive feedback that has greatly helped improve the quality of this work. The TNG50 simulation was run with compute time granted by the Gauss Centre for Supercomputing (GCS) under Large-Scale Projects GCS-DWAR on the GCS share of the supercomputer Hazel Hen at the High Performance Computing Center Stuttgart (HLRS). GCS is the alliance of the three national supercomputing centres HLRS (Universit{\"a}t Stuttgart), JSC (Forschungszentrum J{\"u}lich), and LRZ (Bayerische Akademie der Wissenschaften), funded by the German Federal Ministry of Education and Research (BMBF) and the German State Ministries for Research of Baden-W{\"u}rttemberg (MWK), Bayern (StMWFK) and Nordrhein-Westfalen (MIWF). This analysis has been carried out on the VERA supercomputer of the Max Planck Institute for Astronomy (MPIA), operated by the Max Planck Computational Data Facility (MPCDF). This research has made use of NASA's Astrophysics Data System Bibliographic Services.

\bibliographystyle{mnras}
\bibliography{references}

\end{document}